\documentclass[aps,prd,nofootinbib, twocolumn, superscriptaddress]{revtex4-2}

\usepackage{amsmath,amssymb, amsthm,amstext}
\usepackage{natbib}
\usepackage{graphicx}
\usepackage{color}
\usepackage[dvipsnames]{xcolor}
\usepackage{array, enumerate}
\usepackage{bm}
\usepackage{multirow}
\usepackage[breaklinks,colorlinks,citecolor=cyan,urlcolor=NavyBlue]{hyperref}
\usepackage{braket}
\usepackage{txfonts}
\usepackage{physics}
\usepackage{dcolumn}

\usepackage{tikz}
\usetikzlibrary{decorations.pathmorphing}
\tikzset{snake it/.style={decorate, decoration=snake}}

\def\be{\begin{equation}}
\def\ee{\end{equation}}
\def\ba{\begin{aligned}}
\def\ea{\end{aligned}}
\def\bpm{\begin{pmatrix}}
\def\epm{\end{pmatrix}}

\def\bpsi{{\bar{\psi}}}
\def\bphi{{\bar{\varphi}}}
\def\ta{{\tilde{a}}}

\begin{document}

\title{Black hole superradiance of interacting multifield}
\author{Zhi-Qing Zhu}
\email{zhuzhiqing24@mails.ucas.ac.cn}
\affiliation{International Centre for Theoretical Physics Asia-Pacific, University of Chinese Academy of Sciences, 100190 Beijing, China}
\author{Yun-Song Piao}
\email{yspiao@ucas.ac.cn}
\affiliation{School of Physical Sciences, University of Chinese Academy of Sciences, Beijing 100049, China}
\affiliation{School of Fundamental Physics and Mathematical Sciences, Hangzhou Institute for Advanced Study, University of Chinese Academy of Sciences, Hangzhou 310024, China}
\affiliation{International Centre for Theoretical Physics Asia-Pacific, University of Chinese Academy of Sciences, 100190 Beijing, China}\affiliation{Institute of Theoretical Physics, Chinese Academy of Sciences, P.O. Box 2735, Beijing 100190, China}
\author{Jun Zhang}
\email{zhangjun@ucas.ac.cn}
\affiliation{International Centre for Theoretical Physics Asia-Pacific, University of Chinese Academy of Sciences, 100190 Beijing, China}
\affiliation{Taiji Laboratory for Gravitational Wave Universe, University of Chinese Academy of Sciences, 100049 Beijing, China}

\begin{abstract}
We investigate black hole superradiance evolution of the interacting multiple fields. We consider a model of two scalar fields interacting with a cubic coupling, and study the superradiant evolution of the cloud. We demonstrate that superradiance is typically suppressed when the superradiant field couples to another field, even with a very weak coupling strength. This implies that the constraints on dark particles derived from single-field analyses can be revised in the presence of interactions. Moreover, we find that the multifield superradiant evolution and its corresponding observational signatures can be different across parameter spaces, which makes black hole superradiance an even more powerful probe of the dark sector in particle physics.
\end{abstract}
\maketitle

\section{Introduction}

Bosonic fields around a rapidly rotating black hole manifest an instability known as black hole superradiance~\cite{ZelDovich:1971, Press:1972zz, Zouros:1979iw, Detweiler:1980uk, Brito:2015oca}. In particular, ultralight bosons can extract angular momentum and energy from an astrophysical black hole, forming a massive cloud around it. With recent advances in gravitational wave detection and black hole imaging, black hole superradiance has emerged as a powerful probe of the dark sector in particle physics~\cite{Arvanitaki:2009fg,Arvanitaki:2010sy}. Extensive studies have explored the observational consequences of black hole superradiance, including black hole spin down~\cite{Baryakhtar:2020gao}, gravitational wave emission from the clouds~\cite{Arvanitaki:2014wva,Brito:2017wnc, Brito:2017zvb}, and dynamical signatures of clouds in black hole binaries~\cite{Cardoso:2011xi, Ferreira:2017pth, Baumann:2018vus, Zhang:2018kib, Zhang:2019eid,  Baumann:2019ztm, Berti:2019wnn, Ikeda:2020xvt, Baumann:2021fkf, Baumann:2022pkl, Tong:2022bbl,  Tomaselli:2023ysb, Cao:2023fyv,  Guo:2023lbv, Guo:2024iye, Tomaselli:2024bdd, Cao:2024wby,Takahashi:2024fyq, Tomaselli:2024dbw,Boskovic:2024fga,Guo:2025ckp, Peng:2025zca,Tomaselli:2025jfo}, leading to stringent constraints on ultralight bosonic fields~\cite{Tsukada:2018mbp, Isi:2018pzk, Palomba:2019vxe, Tsukada:2020lgt,Ng:2020ruv, LIGOScientific:2021rnv, Yuan:2022bem, Witte:2024drg, Miller:2025yyx, Mirasola:2025car, Aswathi:2025nxa,Xie:2025npy,Caputo:2025oap}.

While the dark sector may consist of multiple particle species with nontrivial interactions, most studies assume the presence of only a single superradiant field. Although some works have considered interactions between the superradiant field and other fields, their focus has largely been on phenomena, such as photon polarization by the cloud~\cite{Chen:2019fsq, Chen:2021lvo, Chen:2022oad} and annihilation of the cloud into photons~\cite{Spieksma:2023vwl}, fermions~\cite{Chen:2023vkq} and scalars~\cite{Lyu:2025lue}, with the assumption of a preformed superradiant cloud. Not only has the superradiance of the coupled fields (or generally, interactions between the black hole and bound states of the coupled fields) been overlooked, but its consequential effects on the superradiance of the primary field also lack investigation. Therefore, a systematic investigation of superradiant evolution of the interacting multiple fields is essential and necessary.

In this work, we investigate black hole superradiance in the presence of multiple fields. We consider a model of two scalar fields interacting with a cubic coupling, and study the superradiance evolution of the cloud. We demonstrate that superradiance is typically suppressed if the superradiant field couples to another field even with a very weak coupling strength. This implies that the constraints on dark particles derived from single-field analyses can be significantly revised in the presence of interactions. Moreover, the superradiant evolution in different regions of the parameter space are distinct. This makes black hole superradiance an even more powerful probe of the dark sector in particle physics.

The rest of this paper is organized as follows. We start with the superradiance evolution of two noninteracting fields in Sec.~\ref{sec:free}. Then we shall consider a cubic interaction, and shall discuss its effects on the superradiant growth rate in Sec.~\ref{sec:int} and the superradiant evolution in Sec.~\ref{sec:sr}. Finally, we shall discussion the implications observations in Sec.~\ref{sec:obs}. Sec.~\ref{sec:con} is devoted to the conclusion and discussion. We will take $(-,+,+,+)$ metric convention and set $\hbar=c=1$.

\section{Superradiance of free multifields}
\label{sec:free}

In this section, we investigate the superradiance process of noninteracting multifields (also see Ref.~\cite{Neves:2024rju} for a similar study). For the purpose of demonstration, we shall start with a model of two massive scalar fields in the Kerr background, the Lagrangian of which is given by
\be\label{eq:model}
{\cal L}=  -\frac12 g^{ab} \partial_{a} \psi\, \partial_{b} \psi -\frac{1}{2} \mu^2 \psi^2  -\frac12 g^{ab} \partial_{a} \varphi\, \partial_{b} \varphi -\frac{1}{2} \nu ^2 \varphi^2 + {\cal L}_{\rm int} \, ,
\ee
where $g_{ab}$ is the Kerr metric, and $\mu$ and $\nu $ denote the mass of fields $\psi$ and $\varphi$ respectively. Here we have also included ${\cal L}_{\rm int}$, denoting the interactions between the two fields. We shall take ${\cal L}_{\rm int} = 0$ in this section, and will consider the interactions in the later sections. For convenience, we define $\alpha \equiv G M \mu$,  and the mass ratio $q=\nu /\mu$.

\subsection{Eigenstates of free fields}

\begin{figure}
   \includegraphics[height=0.5\linewidth]{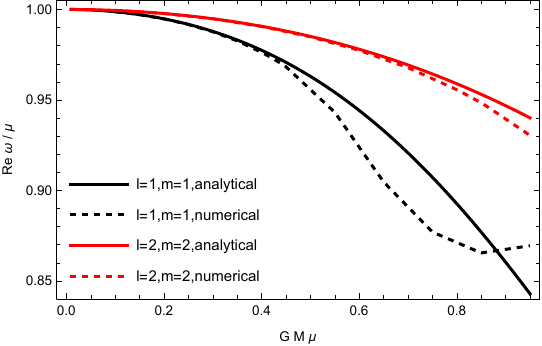} \\
   \includegraphics[height=0.5\linewidth]{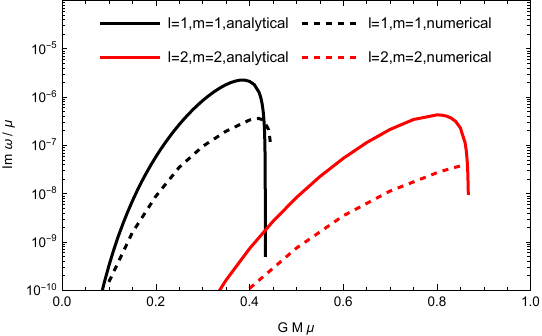}
\caption{Eigenfrequencies of bound states with $n \ell m$ being $211$ (in black)  and $322$ (in red). The upper and lower plots show the real and imaginary parts of the eigenfrequencies respectively. The solid lines are given by Eqs.~\eqref{eq:real_freq} and~\eqref{eq:imag_freq}, while the dashed lines show the results obtained from numerical calculation.}
\label{fig:eigenfrequencies}
\end{figure}

Without direct interactions, the fields only talk to the background, and could be unstable on the Kerr background. The instability manifests in the eigenfrequencies of the fields. Taking $\psi$ for instance, the Klein-Gordon equation satisfied by $\psi$ is separable with the ansatz
\be \label{eq:Psi}
\psi \left(x^{\mu}\right)= e^{- i \omega t} R_{\omega \ell m}(r) S_{\omega \ell m}(\theta) e^{i m \varphi} \, ,
\ee
where $x^\mu = (t, r, \theta, \varphi)$ are the Boyer-Lindquist coordinates, $R_{\omega \ell m}(r)$ is the radial function, and $S_{\omega \ell m}(\theta)$ turns out to be the spheroidal harmonics. The boundary conditions at the black hole horizon and at spatial infinity single out a set of eigenfrequencies. For bound states, i.e., states vanishes at spatial infinity, the eigenfrequencies $\omega$ are shown to be a set of discrete complex number $\omega_{n \ell m}$, labeled by three quantum numbers $n$, $\ell$, and $m$. These eigenfrequencies can be obtained numerically. For instance, Fig.~\ref{fig:eigenfrequencies} shows the eigenfrequencies of some bound states obtained with the continued fraction method~\cite{Leaver:1985ax}. For $\alpha \ll 1$, we have~\cite{Baumann:2018vus}
\be\label{eq:real_freq}
\ba
 {\rm Re}\, \omega_{n \ell m} = \mu \left( 1 -\frac{\alpha^2}{2 n^2}-\frac{\alpha^4}{8 n^4} +\frac{(2\ell-3 n+1)\alpha^4}{n^4(\ell+1/2)}\right. \\
  \left.+\frac{2 \tilde{a} m \alpha^5}{n^3 \ell(\ell+1/2)(\ell+1)} +  {\cal O} (\alpha^{6}) \right) \,,
\ea
\ee
and~\cite{Detweiler:1980uk,Pani:2012bp,Rosa:2012uz}
\be
\label{eq:imag_freq}
  {\rm Im}\, \omega_{n \ell m} \approx  2\tilde{r}_+ (m \Omega_H-\omega_{n \ell m}) \alpha^{4 \ell+5} C_{n \ell m}
\ee
with
\be
\ba
    C_{n \ell m} = \frac{2^{4\ell+2}(2\ell+n+1)!}{(n+\ell+1)^{2\ell+4}n!} \left(\frac{\ell!}{(2\ell)!(2\ell+1)!}\right)^2 \\
    \times \prod^\ell_{j=1} \left( j^2(1-\tilde{a}^2)+(m \tilde{a} - 2 \tilde{r}_+ \alpha)^2 \right) \, ,
     \ea
\ee
where $\tilde{a}$ is the black hole dimensionless spin, $\tilde{r}_+ \equiv 1+\sqrt{1-\tilde{a}^2}$, and $\Omega_H \equiv \tilde{a}/2M\tilde{r}_+$ is the angular velocity of the outer horizon.\footnote{Also see Refs.~\cite{Bao:2022hew,Bao:2023xna} for the improved analytical superradiance solutions, and Refs.~\cite{Dai:2023zcj, Dai:2023ewf} for comparison with Hawking radiation.} Therefore, when $m \Omega_H > \omega_{n \ell m}$, the bound state experiences the superradiant growth with a rate of $\Gamma_{n \ell m} \equiv {\rm Im}\, \omega_{n \ell m}$.
While the wave function of the bound states can also be obtained numerically~\cite{leaver1986solutions}, for $\alpha \ll 1$, it is convenient to work with the wave function in the nonrelativistic limit. Taking the ansatz
\be\label{eq:bpsi}
\psi =  \frac{1}{\sqrt{2 \mu}} \left(\bpsi e^{- i \mu t} + c.c. \right)\,,
\ee
where $\bpsi$ is a complex field that varies on a time scale much longer than $\mu^{-1}$, the Klein-Gordon equation reduces to a Schrödinger-like equation,
\be\label{eq:eombpsi}
\left[i \partial_t + \frac{\nabla^2}{2\mu} +\frac{\alpha}{r} + {\cal O}(\alpha^2) \right]  \bpsi =0 \,,
\ee
and the eigenstates of $\bpsi$ are given by
\be\label{eq:bpsinlm}
\bpsi_{n \ell m} = \sqrt{N_{n\ell m}}  \,u_{n\ell m} ({\bf r})  e^{- i \xi_{n \ell m} t} \, ,
\ee 
where $\xi_{n \ell m} = \omega_{n \ell m} -\mu$, $N_{n\ell m}$ is the occupation number of the particle in the eigenstate, and $u_{n\ell m} ({\bf r})$ is normalized to 1. To the leading order in $\alpha$, $u_{n\ell m} ({\bf r})$ is given by the normalized hydrogenic eigen-wave function with a Bohr radius of $r_B \equiv GM/\alpha^2$. 

The Klein-Gordon equation of $\psi$ also allows unbound states, which satisfy the outgoing boundary condition at spatial infinity. The unbound states are continuous in spectrum, with the eigenfunctions, in the Newtonian limit, given by the stationary Coulomb waves~\cite{landau1977quantum}
\be
\psi_{k \ell m} \propto  u_{k \ell m} ({\bf r}) e^{-i \omega_{k \ell m} t} \, .
\ee
The exact expression of eigenfunctions in Newtonian potential are discussed in Appendix.~\ref{app:weakgravity}.
The above discussion can be easily extend to the $\varphi$ field by replacing $\alpha$ with $q \alpha$. 

\begin{figure}
    \includegraphics[height=0.75\linewidth]{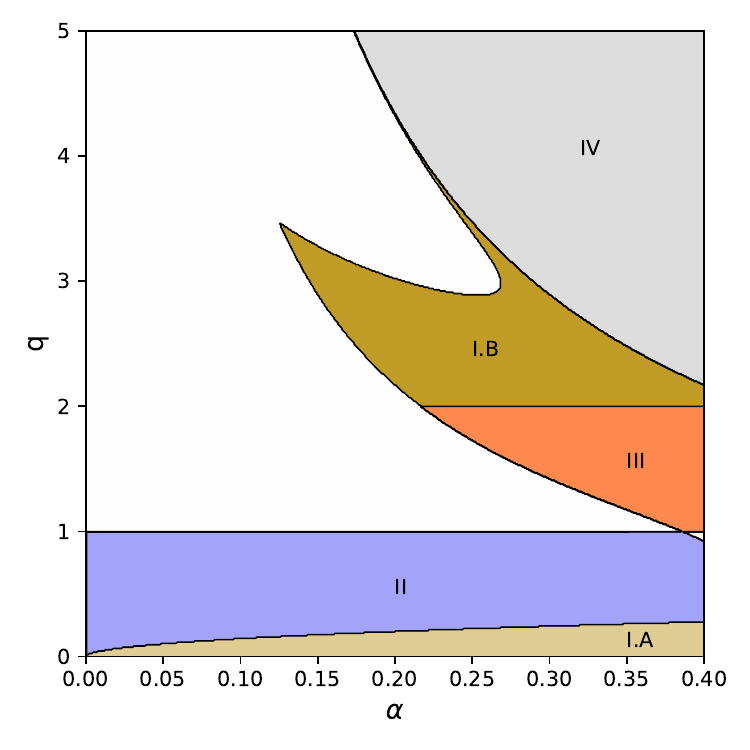}
    \caption{Parameter space of two noninteracting fields. Without loss of generality, we assume $\psi_{211}$ to be the fastest growing mode, which excludes the white regime. Representative examples of the evolution in some regions are shown in Fig.~\ref{fig:ff_evo}, while the boundaries and the superradiant evolution of each region are described in Sec.~\ref{sec:singlese}. In particular, we find that superradiant growth of $\psi_{211}$ might be affected by $\varphi$, which may accelerate black hole spin down, causing $\psi_{211}$ depletes prematurely.}
    \label{fig:ff_para}
\end{figure}

\begin{figure*}
\includegraphics[height=0.2\linewidth]{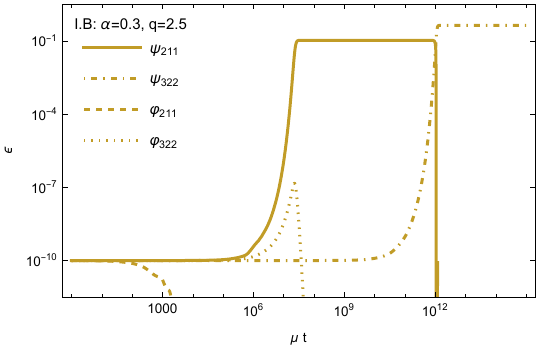}
\includegraphics[height=0.2\linewidth]{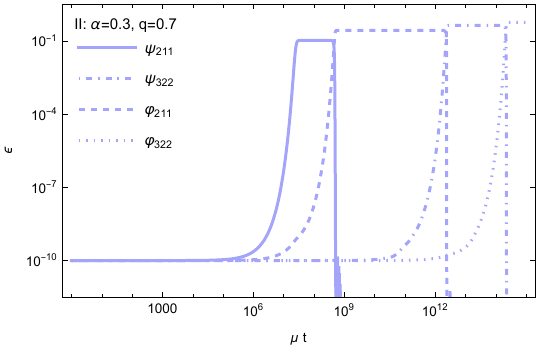}
\includegraphics[height=0.2\linewidth]{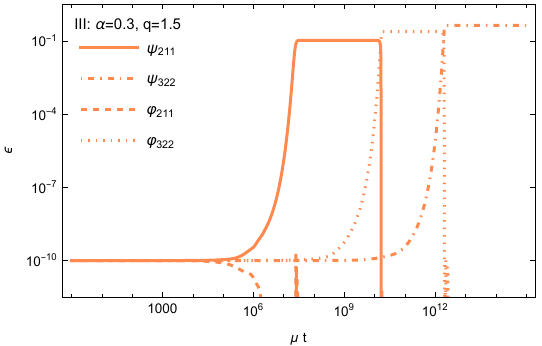} \\
\includegraphics[height=0.2\linewidth]{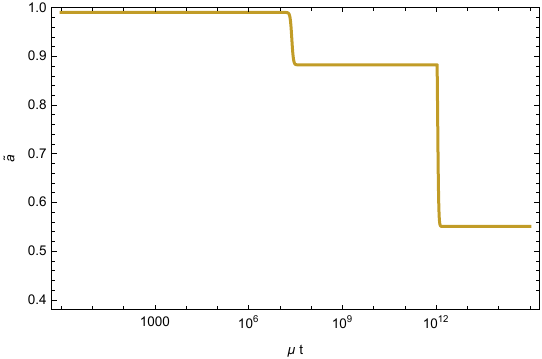}
\includegraphics[height=0.2\linewidth]{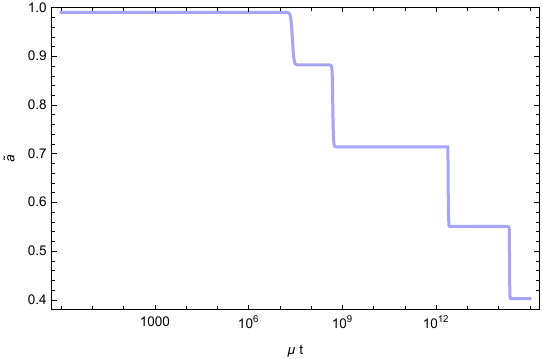}
\includegraphics[height=0.2\linewidth]{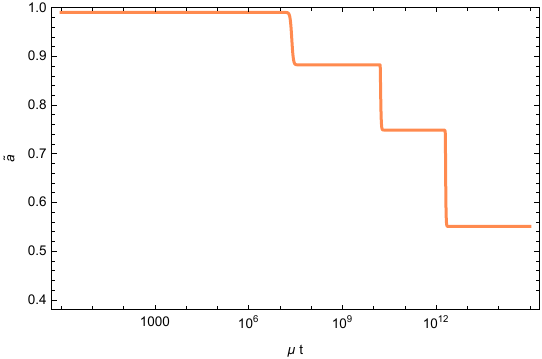}
\caption{Presentative examples of noninteracting two-field superradiant evolution. The upper panel shows the evolution of occupation numbers in the region of corresponding color in Fig.~\ref{fig:ff_para}, while the lower panel shows the evolution of black hole spin. In the left example, $\varphi_{211}$ and $\varphi_{322}$ cannot grow sufficiently because the black hole spin drops quickly due to the superradiant growth of $\psi_{211}$. In the middle example, $\psi_{211}$, $\varphi_{211}$, $\psi_{322}$ and $\varphi_{322}$ undergo efficient superradiant growth one after another and dominate the black hole spin in sequence. In the right example, the superradiant growth of $\varphi_{211}$ is suppressed by that of $\psi_{211}$, while $\varphi_{322}$ and $\psi_{322}$ can still grow in order after $\psi_{211}$ saturates.}
\label{fig:ff_evo}
\end{figure*}

\subsection{Superradiance evolution}
\label{sec:singlese}

The superradiant growth of the fields back reacts on the background geometry, leading to black hole spin down. Under the adiabatic approximation, the angular momentum and mass of the black hole evolve as
\be
\label{sf_da}
    \dot{J} = - GM^2\sum_i m_i \gamma_i \epsilon_i 
\ee
\be\label{sf_dM}    
    \dot{M} = - GM^2 \mu \sum_i \gamma_i \epsilon_i \,,
\ee
where we have defined $\gamma_i = \Gamma_i/\mu$ and $\epsilon_i = N_i/GM^2$, the overdot denotes derivative with respect to the dimensionless time $\tilde{t} \equiv \mu t$, and the summation over all bound states of both $\psi$ and $\varphi$.

In the case of a single filed $\psi$, $\psi_{211}$ typically grows most rapidly with
\be
\gamma_{211} = \frac{1}{24} \alpha^8 \left[ \tilde{a}- 2\alpha \left(1+\sqrt{1-\tilde{a}^2}\right) \right]\,,
\ee
and first dominates the black hole spin evolution. It continues growing until the black hole spin drops below $\tilde{a}_{\rm crit} = 4\alpha/(1+4\alpha^2)$ (for $\alpha < 0.5$), namely the superradiant condition is no longer satisfied. At this time, $\psi_{211}$ saturates with an occupation number of $\epsilon_{max} = \tilde{a}_0-\tilde{a}_{\rm crit} \sim 1$. Similarly, $\psi_{322}$ can also undergo superradiance growth until the black hole drops below $\tilde{a}^{\psi_{322}}_{\rm crit}=2\alpha/(1+\alpha^2)$.

In the case of two fields, the superradiance evolution depends on the parameters. In Fig.~\ref{fig:ff_para}, we show the parameter space of different types of evolution, assuming $\psi_{211}$ to be the fastest growing mode.\footnote{The case that $\varphi_{211}$ grows faster can be obtained by replacing $q$ with $1/q$.} 
Specially, if $\gamma_{\psi_{322}}>\gamma_{\varphi_{211}}$ and $\gamma_{\psi_{322}}>\gamma_{\varphi_{322}}$ (region I.A), $\psi_{322}$ will undergo efficient superradiant growth after $\psi_{211}$ saturates and dominates black hole spin down. This process is then followed by the superradiant growth of $\varphi_{211}$ and $\varphi_{322}$. 
If $\gamma_{\varphi_{211}}<\gamma_{\psi_{322}}<\gamma_{\varphi_{322}}$ and $\tilde{a}^{\varphi_{322}}_{\rm crit}>\tilde{a}^{\psi_{211}}_{\rm crit}$ (region I.B), $\varphi_{211}$ and $\varphi_{322}$ cannot grow sufficiently as the black hole spin drops below their critical spin quickly as $\psi_{211}$ grows, while $\psi_{322}$ can still grow sufficiently after $\psi_{211}$ saturates.
If $\gamma_{\varphi_{211}}>\gamma_{\psi_{322}},\gamma_{\varphi_{322}}$ and $\tilde{a}^{\varphi_{211}}_{\rm crit}<\tilde{a}^{\psi_{211}}_{\rm crit}$ (region II), $\varphi_{211}$ will undergo efficient superradiant growth after $\psi_{211}$ saturates and dominates black hole spin down. This process is then followed by the superradiant growth of $\psi_{322}$ and $\varphi_{322}$. 
If $\gamma_{\varphi_{322}}>\gamma_{\psi_{322}}$ and $\tilde{a}^{\varphi_{211}}_{\rm crit}>\tilde{a}^{\psi_{211}}_{\rm crit}, \tilde{a}^{\varphi_{322}}_{\rm crit}<\tilde{a}^{\psi_{211}}_{\rm crit}$ (region III), $\varphi_{211}$ will not grow, while $\varphi_{322}$ will grow efficiently after $\psi_{211}$ saturates, followed by the superradiant growth of $\psi_{322}$.
A similar analysis can be extended in region IV, where higher modes of $\varphi$, such as $\varphi_{433}$, should also be taken into account. A representative example of each type is shown in Fig.~\ref{fig:ff_evo}. These examples are obtained by solving Eq.~\eqref{sf_da} together with $\dot{\epsilon}_i = \gamma_i \epsilon_{i}$, assuming the initial conditions of $\epsilon_i = 10^{-10}$ when the black hole spin $\tilde{a}_0 = 0.99$. For simple, we treated the black hole mass (and hence $\alpha$) to be constant, given the fact that the black hole mass typically varies by a small amount during superradiant evolution. We find that, in region I and region IV, $\psi_{211}$ finally depletes after reaching the maximum occupation number, due to the development of the $\varphi$ field.

\section{Cubic Interactions}
\label{sec:int}

In this section, we investigate the superradiant growth of interacting scalar fields. We shall consider cubic interactions which typically dominate over other higher dimensional interactions. We shall focus on the interactions between the fields instead of self-interactions, as those have been investigated in Ref.~\cite{Baryakhtar:2020gao}. This leaves two relevant operators: $\psi^2 \, \varphi$ and $\psi \, \varphi^2$. As the growth rate is extremely sensitive to the mass of the fields, one of the fields usually grow much faster than another field, even if the masses of the two field are comparable. Without loss of generality, let $\psi$ be the one that grows faster, in which case $\psi^2 \, \varphi$ is expected to dominate over $\psi \, \varphi^2$. For these reasons, we shall neglect $\psi \, \varphi^2$, and consider Lagrangian~\eqref{eq:model} with ${\cal L}_{\rm int} = \lambda \psi^2 \, \varphi$, where $\lambda$ is the coupling constant. 

The field equations are
\begin{align}
&\left(\Box - \mu^2\right) \psi = -2 \lambda \psi \varphi \,, \label{eq:eompsi} \\
&\left(\Box - \nu ^2\right) \varphi = -\lambda \psi^2 \,, \label{eq:eomphi}
\end{align}
where $\Box$ denotes the D'Alembert operator in the Kerr background. We shall further assume the two fields are weakly coupled.

Assuming both fields are initially in vacuum states with small quantum fluctuations, the superradiance process can be investigated perturbatively: At the very beginning, interactions between the field are suppressed due to the low occupation numbers, and the superradiance process is the same as it in the case of free field. As superradiance continues, the occupation number in the fastest growing mode, i.e., $\bpsi_{211}$, increases, resulting in a notable interaction between the two fields. Depending on the parameters, the interaction may lead to different processes: The $\psi$ field may lose energy to the black hole via the bound state of $\varphi$. It may also lose energy to the spatial infinity via the unbound states of $\varphi$ and $\psi$. The effects on the superradiant growth of such processes can be investigated by their corrections on the eigenfrequencies, which will be discussed in what follows.

\subsection{Bound state interactions}

We shall start with interactions between $\varphi$ and the bound states of $\psi$. It is convenient to described the bound states with $\bpsi$ defined in the nonrelativistic ansatz~\eqref{eq:bpsi}, and work in the weak gravity limit. By substituting ansatz~\eqref{eq:bpsi} and expanding terms in powers of $\alpha$, Eqs.~\eqref{eq:eompsi} and \eqref{eq:eomphi} can be written into the form of
\be
\ba\label{eq:eomm}
&\left( i \partial_t  - \hat{H} \right) \left( \bpsi e^{-i \mu t} + \bpsi^* e^{i \mu t} \right)= -2 \bar{\lambda} \left( \bpsi e^{-i \mu t} + \bpsi^* e^{i \mu t} \right) \varphi  \, , \\ 
&\left(\Box - \nu^2 \right) \varphi = \bar{\lambda} \left( \bpsi^2 e^{-2 i \mu t} + 2\bpsi \bpsi^* + {\bpsi}^{*2} e^{2 i \mu t} \right),
\ea
\ee
where $\bar{\lambda} \equiv \lambda/2 \mu$ is the reduced coupling strength, and $\hat{H} \simeq - \tfrac{\nabla^2}{2\mu} - \tfrac{\alpha}{r} $ to the leading order of $\alpha$ (see Appendix.~\ref{app:weakgravity} for the detailed derivation). Here we do not take the nonrelativistic ansatz for $\varphi$, as its relativistic states could also contribute in this process.

To compute the corrections on the eigenfrequency, we take one of the $\bpsi$ in the source terms as background $\bpsi_0 $. In our case, the background is mostly in the $\bpsi_{211}$ mode, i.e., $\bpsi_0 = \sqrt{N_0} u_{211} (\mathbf{r}) e^{-i \xi_{211} t}$. While $\bpsi_{211}$ should, in principle, grow with time, given the fact that ${\rm Im} \xi_{211} \ll {\rm Re} \xi_{211}$, one can neglect the imaginary part of $\xi_{211}$ and treat  $\bpsi_0$ to be stationary in our perturbative calculation for simplicity~\cite{Baryakhtar:2020gao}. It is convenient to use Feynman-like diagrams to represent the channels involved in Eq.~\eqref{eq:eomm}. In particular, $\bpsi$ on the rhs of Eq.~\eqref{eq:eomm} corresponds to a leg on the lhs of the vertex, and vice versa; and $\bpsi^*$ on the rhs of Eq.~\eqref{eq:eomm} corresponds to a leg on the rhs of the vertex, and vice versa. We are interested in the interaction corrections on the fastest growing mode $\bpsi_{211}$. Given Eq.~\eqref{eq:eomm}, there are three possible channels, which are shown in Fig.~\ref{fig:channels}.

\begin{figure*}[t]
\centering
    \begin{minipage}{0.23\textwidth}
        \centering
        \begin{tikzpicture}
        \draw (-1,1) -- (0,0);
        \draw (-1,-1) -- (0,0);
        \draw [dashed] (0,0) -- (1,0);
        \draw (2,1) -- (1,0);
        \draw (2,-1) -- (1,0);
        \node[anchor=south] at (-1,1) {$\psi_{211}$};
        \node[anchor=north] at (-1,-1) {$\psi_{211}$};
        \node[anchor=south] at (2,1) {$\psi_{211}$};
        \node[anchor=north] at (2,-1) {$\psi_{211}$};
        \node[anchor=north] at (0,-2) {${}$};
        \end{tikzpicture}
    \end{minipage}
    \hfill
    \begin{minipage}{0.23\textwidth}
        \centering
        \begin{tikzpicture}
        \draw (-1,1) -- (0,0);
        \draw (1,1) -- (0,0);
        \draw [dashed] (0,0) -- (0,-1);
        \draw (0,-1) -- (1,-2);
        \draw (0,-1) -- (-1,-2);
        \node[anchor=south] at (-1,1) {$\psi_{211}$};
        \node[anchor=south] at (1,1) {$\psi_{211}$};
        \node[anchor=north] at (1,-2) {$\psi_{211}$};
        \node[anchor=north] at (-1,-2) {$\psi_{211}$};
        \end{tikzpicture}
    \end{minipage}
    \hfill
        \begin{minipage}{0.23\textwidth}
        \centering
        \begin{tikzpicture}
        \draw (-1,1) -- (0,0);
        \draw (0,0) -- (1,-2);
        \draw [dashed] (0,0) -- (0,-1);
        \draw (0,-1) -- (1,1);
        \draw (0,-1) -- (-1,-2);
        \node[anchor=south] at (-1,1) {$\psi_{211}$};
        \node[anchor=south] at (1,1) {$\psi_{211}$};
        \node[anchor=north] at (1,-2) {$\psi_{211}$};
        \node[anchor=north] at (-1,-2) {$\psi_{211}$};
        \end{tikzpicture}
    \end{minipage}
        \caption{Possible channels involved in Eq.~\eqref{eq:eomm}. The left diagram corresponds to the $s$ channel described by Eqs.~\eqref{eq:scpsit} and \eqref{eq:scphit}, while the middle and right diagrams correspond to the $t$ and $u$ channels described by Eq.~\eqref{tu_channel}.}
    \label{fig:channels}
\end{figure*}

\begin{figure*}
\includegraphics[height=0.2\linewidth]{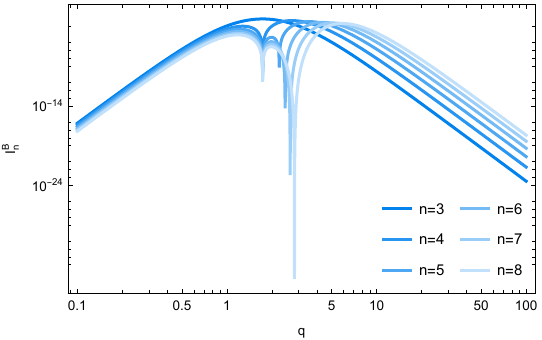} 
\includegraphics[height=0.2\linewidth]{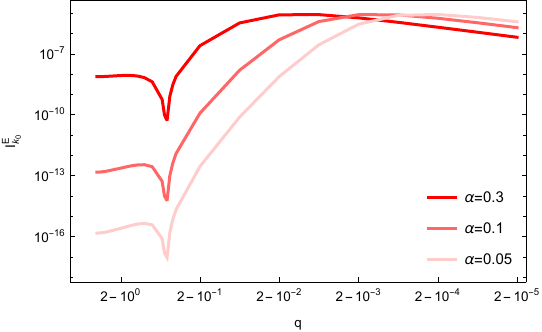} 
\includegraphics[height=0.2\linewidth]{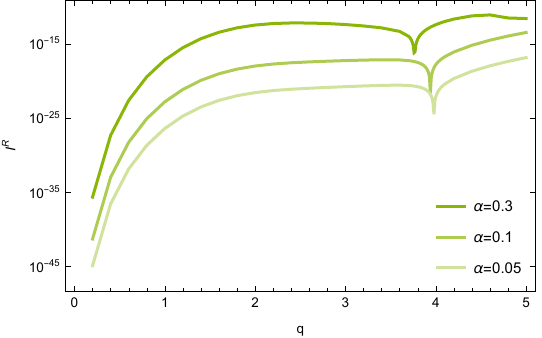} 
\caption{Numerical results of integrals \eqref{Ib},~\eqref{Ie}, and~\eqref{Ir}. In practice, we have to impose a cutoff $n_{\rm max}$ for the summation in Eq.~\eqref{dwB}. Given the left plot, we expect the summation converges well when $n_{\rm max} > 8$ for $q<5$.}
\label{fig:integral}
\end{figure*}

For the $s$ channel, we have
\begin{align}
&\left( i \partial_t - \hat{H} \right) \bpsi = -2 \bar{\lambda} \bpsi_0^* \varphi e^{2 i \mu t} \,, \label{eq:scpsit}\\
&\left( \Box - \nu^2 \right) \varphi = \bar{\lambda} \bpsi \bpsi_0 e^{-2 i \mu t} \,.\label{eq:scphit}
\end{align}
To investigate the interaction corrections on $\omega_{211}$, we shall consider
\be
\bpsi = \bpsi^{(0)} + \bpsi^{(1)} + \cdots
\ee
with $\bpsi^{(0)} = u_{211} e^{- i \xi_{211} t}$ and solve Eqs.~\eqref{eq:scpsit} and~\eqref{eq:scphit} perturbatively. Here the superscription denotes the order of the perturbation calculation.

We shall start with Eq.~\eqref{eq:scphit}. To the leading order in $\bar{\lambda}$, the rhs of Eq.~\eqref{eq:scphit} is given by $\bar{\lambda} \bpsi^{(0)} \bpsi_0 e^{-2i\mu t}$. Then by considering the ansatz
\be\label{eq:sourcephi}
\varphi(t,\mathbf{r}) =\Phi(\mathbf{r})\, e^{-i \sigma_0 t} \,,
\ee
Eq.~\eqref{eq:scphit} leads to $\sigma_0 = 2\omega_{211}$ and
\be
\left(\hat{M}_{\sigma = \sigma_0} - \nu^2 \right) \Phi = \tilde{s}\left(\mathbf{r}\right) \, ,
\ee
with $\hat{M}_{\sigma}$ can be obtained by replacing $\partial_t$ in $\Box$ with $- i \sigma$ and $\tilde{s}\left(\mathbf{r}\right) = \bar{\lambda}\sqrt{N_0} u_{211}^2$. The above equation can be solved with the Green's function method,\footnote{Technically, the integral should be performed with the tortoise coordinates $\mathbf{r}_*$. See Appendix.~\ref{app:Green} for the precise formula. Nevertheless, we shall not distinguish $\mathbf{r}_*$ with $\mathbf{r}$ here, as $\tilde{s}(\mathbf{r})$ only has support around $r \sim GM/\alpha^2$ where $\mathbf{r}_* \simeq \mathbf{r}$.}
\be
\Phi (\mathbf{r}) = \int {\rm d}^3 \mathbf{r} \,G_{\sigma=\sigma_0} ( \mathbf{r}, \mathbf{r}') \, \tilde{s}\left( \mathbf{r}'\right)\,.
\ee
In Appendix.~\ref{app:Green}, we show that the above integral can be written as
\be
\ba\label{eq:expPhi}
&\Phi (\mathbf{r}) = \sum_{n, \ell, m}^{\infty} b_{n\ell m} v_{n \ell m}(\mathbf{r}) +  \sum_{\ell, m} \int_{0}^{\infty} \mathrm{d}k \, b_{\ell m}(k) v_{k \ell m} (\mathbf{r})  \,,
\ea
\ee
where $v_{n \ell m}$ and $v_{k \ell m}$ satisfying
\be
\left(\hat{M}_{\sigma = \sigma_{n\ell m}}- \nu^2\right) v_{n \ell m} = 0\,, \quad \left(\hat{M}_{\sigma = \sigma_{k\ell m}}- \nu^2\right) v_{k \ell m} = 0 \,,
\ee
are the bound and scattering state wave functions of the free $\varphi$ field. Here the bound states have ${\rm Re} \sigma_{n\ell m} < \nu$ and are discrete in spectrum, while the scattering states have ${\rm Re} \sigma_{k\ell m} \ge \nu$ and are continuous in spectrum.\footnote{Also see Ref.~\cite{Fu:2025ztk} for a recent discussion on the complete set of mode functions.} The precise formulas for the coefficients $b_{n\ell m}$ and $b(k)_{\ell m}$ are derived in Appendix.~\ref{app:Green}. In practice, when we compute the numerical value of the coefficients, we approximate the wave functions $v_{n \ell m}$ and $v_{k \ell m}$ with their analytical expressions in the weak gravity limit (discussed in Appendix.~\ref{app:weakgravity}), since $\tilde{s} (\mathbf{r})$ only has support far from the black hole horizon. It turns out that the coefficients are nontrivial only if $\ell = m = 2$ due to the angular dependence, and are given by
\be
\ba
b_{n} &\simeq \frac{\bar{\lambda}\sqrt{N_0}}{\sigma_{n\ell m}^2-\sigma_0^2} \int   u_0^2 v_{n 22}^* \,\mathrm{d}^3\mathbf{r}\\
b(k) & \simeq \frac{1}{2\pi}\frac{\bar{\lambda}\sqrt{N_0}}{\sigma_{k\ell m}^2-\sigma_0^2+} \int u_0^2 v_{k 22}^* \,\mathrm{d}^3\mathbf{r} 
\ea
\ee
where we have omitted the subscriptions ${\ell m}$ for short.

With the solution of $\varphi$, we now proceed to Eq.~\eqref{eq:scpsit}. Without lose of generality, we write
\be\label{eq:bpsi1}
\bpsi^{(1)} = u_{211}(\mathbf{r}) e^{- i \delta \omega_{211}^s t} + \sum_{n\ell m \neq 211} a_{n\ell m} \, u_{n \ell m}(\mathbf{r}) e^{- i \omega_{211} t} \,,
\ee
where $u_{n \ell m}$ denotes the bound states of $\bpsi$, and the summation excludes $u_{211}$ as which can be absorbed into $\bpsi^{(0)}$. Substituting Eq.~\eqref{eq:bpsi1} into Eq.~\eqref{eq:scpsit}, and using the orthogonality of $u_{n \ell m}$, we can find $a_{n\ell m} = 0$ and the frequency correction from the $s$ channel $\delta \omega_{211}^{s}$. Specifically, the correction can be written as $\delta \omega_{211}^{s} = \delta \omega_{211}^{sB} + \delta \omega_{211}^{sE}$ with
\be
\ba
&\delta \omega_{211}^{sB} = 2 \bar{\lambda}^2 N_0  \sum_{n \ge 3} \frac{\left| \int u_0^2 v_{n 22}^* \mathrm{d}^3\mathbf{r} \right|^2}{k_0^2+\kappa_n^2} \\
& \delta \omega_{211}^{sE} = 2 \bar{\lambda}^2 N_0  \int_0^\infty \frac{\mathrm{d}k}{2\pi}\, \frac{\left| \int u_0^2 v_{k 22}^* \mathrm{d}^3\mathbf{r} \right|^2}{k_0^2 - \kappa^2(k)}  \,,
\ea
\ee
representing the corrections from bound states and unbound states of $\varphi$ respectively. Here, we have defined $k_0^2\equiv 4\omega_0^2-\nu^2$, and $k \equiv \mathrm{Re}\, \kappa$.

We are interested in the interaction corrections on the growth rate, which can be inferred from the imaginary part of $\delta \omega_{211}^{s}$. For the corrections from bound states, we find
\be\label{dwB}
\ba
&{\rm Im} \delta \omega_{211}^{sB} \\
=& 2 \bar{\lambda}^2 N_0  \sum_{n \ge 3} {\rm Im} \left( \frac{1}{k_0^2+\kappa_n^2} \right) \left| \int u_0^2 v_{n 22}^* \mathrm{d}^3\mathbf{r} \right|^2 \\
\simeq & 2 \bar{\lambda}^2 N_0 \frac{2 q^2 \alpha^3 \mu}{(q^2-4)^2} \sum_{n \ge 3} \frac{ {\rm Im} \sigma_n}{\nu} {\rm I}_n^{B} (q) \,,
\ea
\ee
which indicates that the superradiant growth rate of $\bpsi_{211}$ is suppressed if it interacts with a decaying $\bphi_{n22}$, as $\bphi_{n22}$ may lose energy to the black hole in the $s$ channel, and vice versa.
In the last line of Eq.~\eqref{dwB}, we considered that $\omega_0 \simeq \mu$ and ${\rm Re} \sigma_n \simeq q \mu$, and have defined 
\be\label{Ib}
{\rm I}_n^{B} \equiv \frac{1}{\mu^3\alpha^3} \left| \int u_0^2 v_{n 22}^* \mathrm{d}^3\mathbf{r} \right|^2 \, ,
\ee
which is a dimensionless factor depending only on the mass ratio $q$, with its values shown in Fig.~\ref{fig:integral}. For the corrections from unbound states, we find that $\delta \omega_{211}^{sE}$ has a nonzero imaginary part only if $k_0$ is real due to the optical theorem. Especially, by performing the integral over $k$ with residue theorem, one has
\be
{\rm Im} \delta \omega_{211}^{sE} \simeq  \left\{
                \begin{array}{ll}
              -   \bar{\lambda}^2 N_0   \frac{\mu \alpha^2 }{2 \sqrt{4 -q^2} } {\rm I}^{E}_{k_0}(q,\alpha), \quad &2 \omega_0 > \nu \\
                0, \quad &2 \omega_0 < \nu
                \end{array}
                \right. \, ,
\ee
which represents the superradiant growth of $\bpsi_{211}$ suppressed by their annihilation into radiation of $\varphi$, a process happening only if $2 \omega_0 \ge \nu$. Here
\be\label{Ie}
{\rm I}^{E}_{k_0} \equiv \frac{1}{\mu^2\alpha^2} \left| \int u_0^2 v_{k_0 22}^* \mathrm{d}^3\mathbf{r} \right|^2
\ee
is a dimensionless factor, with its values shown in Fig.~\ref{fig:integral}. 

One should notice that the above calculation is valid only if the interaction terms are perturbative, in other words, when $|\mathrm{Re}\, \delta \omega |\ll |\xi_{211}|$. Since $\mathrm{Re}\, \delta \omega$ is dominated by $\delta \omega^{sB}$, we can define a threshold particle number
\be
N_{\rm th} \equiv \left| \frac{q^2-4}{2 \bar{\lambda}^2 \alpha \sum_n I_n^B} \right|
\ee
beyond which the interaction effects cannot be treated perturbatively. In addition, there are $t$ and $u$ channels which are described by
\be\label{tu_channel}
\ba
&\left( i \partial_t - \hat{H} \right) \bpsi = -2 \bar{\lambda} \bpsi_0 \varphi \, , \\
&\left( \Box - \nu^2 \right) \varphi = \bar{\lambda} \left( \bpsi \bpsi_0^* + \bpsi^* \bpsi_0 \right) \,,
\ea
\ee
and can be investigated in a similar way. The frequency corrections, however, happen to be real, and therefore will not be discussed in detail.

\subsection{Interactions between bound states and unbound states of $\psi$}

Besides radiating $\varphi$, the superradiant mode could also lose energy to spacial infinity by radiating $\psi$ via a process depicted by the right diagram in Fig.~\ref{fig:channel3}. This process, hereinafter referred to as the $3\bpsi$ process, is described by
\begin{align}
&\left(\Box - \mu^2\right) \psi = - 2\lambda \sqrt{\frac{N_0}{2\mu}} u_0 e^{-i \omega_0 t} \bphi + c.c.\,, \label{eq:eompsir} \\
&\left(\Box - \nu ^2\right) \varphi =  - \bar{\lambda} N_0 u_0^2 e^{-2i \omega_0 t} \,, \label{eq:eomphir}
\end{align}
which are obtained from Eqs.~\eqref{eq:eompsi} and~\eqref{eq:eomphi}, with $\psi$ on the rhs being the background $\bpsi_0$. Here we do not take the nonrelativistic ansatz for $\psi$ because the radiated $\psi$ is unbounded and can be relativistic. The effects of the $3\bpsi$ process on the growth rate of $\bpsi_{211}$ can be investigated by estimating the radiation power of $\psi$. 

\begin{figure}[t]
        \begin{minipage}{0.23\textwidth}
        \centering
        \begin{tikzpicture}
        \draw (-1,1) -- (0,0);
        \draw (-1,-1) -- (0,0);
        \draw [dashed] (0,0) -- (1,-1);
        \draw (0,-2) -- (1,-1);
        \draw (1,-1) -- (2,-1);
        \node[anchor=south] at (-1,1) {$\psi_{211}$};
        \node[anchor=north] at (-1,-1) {$\psi_{211}$};
        \node[anchor=north] at (0,-2) {$\psi_{211}$};
        \node[anchor=south] at (2,-1) {$\psi_{\infty}$};
        \end{tikzpicture}
    \end{minipage}
        \caption{The $3\bpsi$ channel that corresponds to Eqs.~\eqref{eq:eompsir} and \eqref{eq:eomphir} .}
    \label{fig:channel3}
\end{figure}

\begin{figure*}
    \centering
    \includegraphics[width=0.3\linewidth]{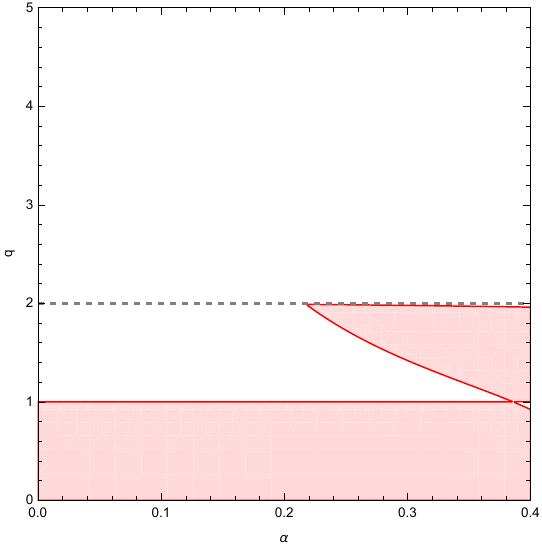}
    \includegraphics[width=0.3\linewidth]{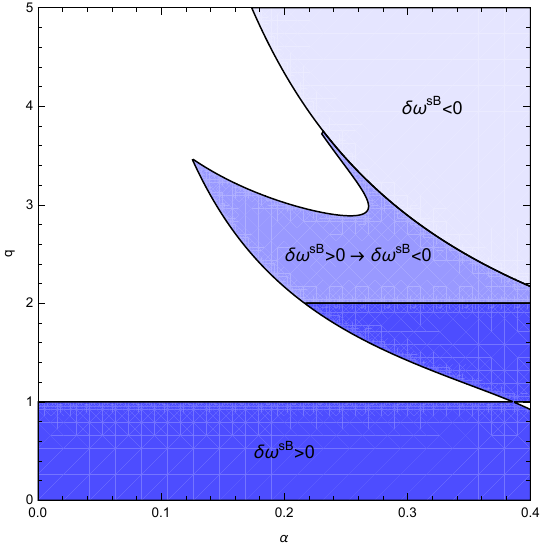}
     \includegraphics[width=0.3\linewidth]{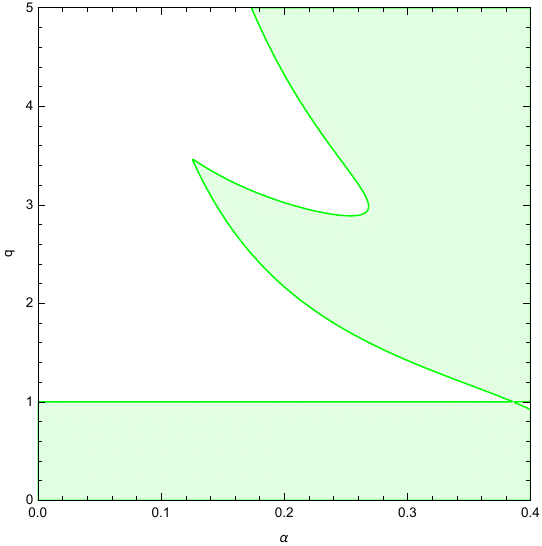}
    \caption{Parameter space of interaction corrections. From left to right, the plots show the regions of nontrivial $\delta \omega_{211}^{sE}$ (in red), $\delta \omega_{211}^{sB}$ (in blue), and $\Gamma_{R}$ (in green). Again, by assuming $\bpsi_{211}$ to be the fastest growing mode, we first exclude the white regime in the middle and right plots. For $\delta \omega_{211}^{sE}$ to be nontrivial, it further requires $q \lesssim 2$. Moreover, we distinguish the regions that ${\rm Im}\delta \omega_{211}^{sB} > 0$ (in dark blue), ${\rm Im}\delta \omega_{211}^{sB} < 0$ (in light blue), and ${\rm Im}\delta \omega_{211}^{sB} $ changes from positive to negative as the black hole spin evolves from $\sim 1$ to $a_{crit}$ (in blue).
}
    \label{fig:para}
\end{figure*}

Again, $\varphi$ can be solved from Eq.~\eqref{eq:eomphir} with the Green's function method,
\be\label{eq:phisln}
 \varphi  = \left( \sum_{n \ge 3} c_{n} \, v_{n22} + \int {\rm d} k\, c(k) \, v_{k 22} \right) e^{-2i \omega_0 t} \,,
\ee
with
\be\label{eq:coe}
\ba
&c_n = -\frac{\bar{\lambda}N_0}{k_0^2+k_n^2} \int u_0^2 v_{n 22}^* \,\mathrm{d}^3\mathbf{r} \\ &c(k) = \frac{1}{2\pi} \,\frac{\bar{\lambda}N_0}{k^2-k_0^2} \int u_0^2 v_{k 22}^* \,\mathrm{d}^3\mathbf{r} \,,
\ea
\ee
where we have considered the fact that only modes with $\ell = m =2$ contribute. Substituting Eq.~\eqref{eq:phisln} into Eq.~\eqref{eq:eompsir} leads to
\be
\left(\Box - \mu^2\right) \psi = f(\mathbf{r})e^{-i \omega_R t}+c.c.
\ee
where $\omega_{R}=3\omega_0$ is the driving frequency, and 
\be
f(\mathbf{r}) \equiv   - 2\lambda \sqrt{\frac{N_0}{2\mu}} u_0 \left( \sum_n c_{n} \, v_{n 22} + \int {\rm d} k\, c(k) \, v_{k 22} \right) \,.
\ee
For simplicity, we solve the above equation using the flat-space approximation. Then the radiation power of $\psi$ is given by
\be
\ba
P = \int \mathrm{d} \Omega_k \frac{2 \omega k}{(4\pi)^2} \abs{\tilde{f}(\mathbf{k})}^2 \,,
\ea
\ee
where $\mathrm{d}\Omega_k$ is the differential solid angle pointing the direction $(\theta_k,\phi_k)$, and 
\be\label{eq:tf}
 \tilde{f}(\mathbf{k}) = 4\pi \sum_{\ell m} Y_{\ell m}(\theta_k,\phi_k) \int \mathrm{d}^3 \mathbf{r}\, (-i)^{\ell} f(\mathbf{r}) \frac{u^*_{k \ell m}}{2k} \, .
\ee
The angular integral in Eq.~\eqref{eq:tf} indicates that only $u^*_{k \ell m}$ with $\ell = m =3$ contributes, and eventually the radiation power can be written as
\be
\ba
P = \frac{4 \omega_R \mu^2}{\sqrt{\omega_R^2-\mu^2}} N_0^3 \bar{\lambda}^4 \alpha^5 {\rm I}^R(\alpha,q) \,,
\ea
\ee
where
\be\label{Ir}
\ba
{\rm I}^R(\alpha,q) \equiv \frac{1}{\alpha^5 \mu} \left| \sum_n \frac{\int u_0^2 v_{n22}^* \mathrm{d}^3 \mathbf{r}}{-k_n^2-k_0^2} \int u_0 v_{n22} u_{k \ell m}^* \mathrm{d}^3 \mathbf{r} \right. \\
\left.+ \int \frac{\mathrm{d}k}{2\pi} \frac{\int u_0^2 v_{k22}^* \mathrm{d}^3 \mathbf{r}}{k^2-k_0^2} \int u_0 v_{k22} u_{k \ell m}^* \mathrm{d}^3 \mathbf{r} \right|^2
\ea
\ee
is a dimensionless quantity including all summations and integrals. The numerical value of some ${\rm I}^R(\alpha,q)$ is shown in Fig.~\ref{fig:integral}. Because of energy conservation, the radiation turns on only if $3\omega_0 \ge \mu$ which is always satisfied providing $\psi$ is a superradiant field. Given the energy of each radiated particle is of $\omega_R=3\omega_0$, we can further define a decay rate given the radiation power, 
\be
\Gamma_R \equiv \frac{P}{\omega_R N_0} \simeq \sqrt{2} \mu N_0^2 \bar{\lambda}^4 \alpha^5 {\rm I}^R(\alpha,q)\,.
\ee
The parameter space for the processes discussed above are summarized in Fig.~\ref{fig:para}.

\begin{figure}
    \centering
     \includegraphics[width=0.75\linewidth]{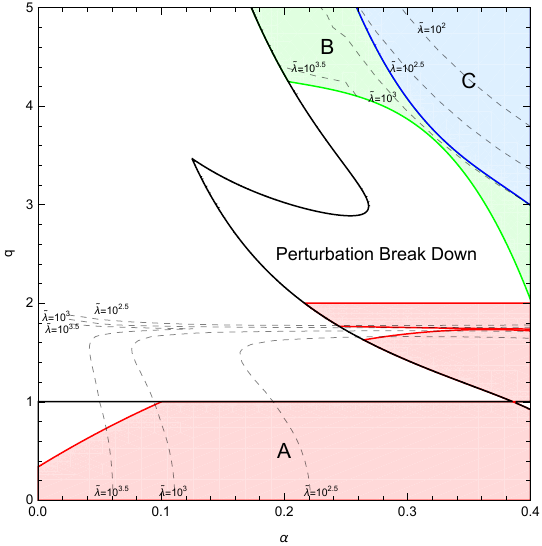}
    \caption{Parameter space of the dominating interaction corrections. We show the regions that the term of $\gamma_E$ (in red), $\gamma_B$ (in blue), and $\gamma_R$ (in green) might dominate the superradiant evolution. We further exclude the region that $\epsilon_{eq} > \epsilon_{th} \equiv N_{th}/GM^2$, cf. Fig.~\ref{fig:para}, where the perturbative treatment may break down. Moreover, we also show the contours of $\tilde{\lambda}_c$ in each color regions. We expect that the interaction correction can interrupt the superradiant growth of $\psi_{211}$ significantly, and lead to the quasiequilibrium stage if $\tilde{\lambda} > \tilde{\lambda}_c$. Given the value of $\tilde{\lambda}_c$, we find that the superradiant growth of $\bpsi_{211}$ can be easily affected by $\varphi$ even with very weak coupling, i.e., $\lambda /\mu \ll 1$.
}
    \label{fig:parad}
\end{figure}

\section{Superradiance Evolution of interacting fields}
\label{sec:sr}

\begin{figure}
\includegraphics[height=0.53\linewidth]{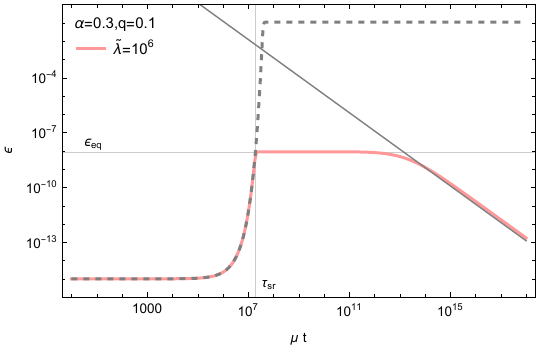}\\
\includegraphics[height=0.5\linewidth]{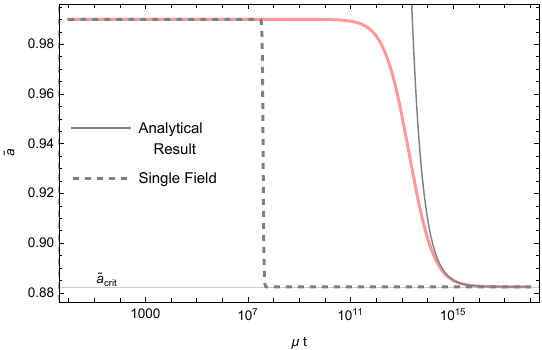}
\caption{The three stages of interacting two-field superradiant evolution. The left and right plots show the evolution of the dimensionless occupation number $\epsilon_{211}$ and the black hole spin respectively, where the vertical grids denotes $\tau_{sr}$ defined in Eq.~\eqref{tausr}, while the horizontal grids denote $\epsilon_{eq}$ and $a_{crit}$. The gray oblique line in the left plot are given by Eq.~\eqref{asymp}, showing the asymptotical behavior.}
\label{fig:stage}
\end{figure}

In this section, we discuss superradiance evolution of the $\bpsi_{211}$ mode in the presence of the interaction. We shall neglect the effects from the superradiant growth of the other modes, and will justify this treatment later. Given the possible processes discussed in Sec.~\ref{sec:int}, the superradiance evolution can be understood by considering the evolution of the occupation number in $\bpsi_{211}$
\be\label{nde}
\dot{ \epsilon}_{211} = \gamma_{211} \epsilon_{211} - \left( \gamma_B + \gamma_E \right) \epsilon_{211}^2 - \gamma_R \epsilon_{211}^3 
\ee
and its effects on the black hole spin
\be\label{nda}
\dot{\tilde{a}} = -\gamma_{211} \epsilon_{211} + \gamma_B \epsilon_{211}^2 
\ee
where the over-dot denotes the derivative with respect to $\tilde{t} = \mu t$, and we have defined
\be
\ba
\gamma_{B} = -\frac{2{\rm Im} \delta \omega^{sB}_{211}}{\epsilon_{211}\, \mu}, \quad \gamma_{E} = - \frac{2{\rm Im} \delta \omega^{sE}_{211}}{\epsilon_{211}\, \mu}, \quad \text{and} \quad \gamma_{R} = \frac{3\Gamma_{R}}{\epsilon_{211}^2\, \mu} \,.
\ea
\ee
In principle, the black hole mass also evolves during the superradiance evolution. Nevertheless, we treat the black hole mass to be constant, since the black hole mass typically varies by a small amount during superradiance. This is especially the case in the presence of the interaction, as the cloud tends to grow to a smaller occupation number. With this treatment, $\alpha$ can also be considered as a constant. We could also take into account gravitational wave radiations by appending $-\gamma_{GW} \epsilon_{211}^2$ on the rhs of both Eq.~\eqref{nde}  with $\gamma_{GW} = \alpha^{14}/160$~\cite{Brito:2014wla,Yoshino:2015nsa}. While we are interested in the case where $\epsilon_{eq}$ derived from the interaction corrections is less than 1, the gravitational wave radiations lead to an effective $\epsilon_{eq} \gg 1$ due to the weak nature of gravitational interactions, and therefore is irrelevant for the discussion.
Figure.~\ref{fig:stage} shows a typical evolution in the presence of interaction, which is obtained by solving Eqs.~\eqref{nde} and~\eqref{nda} numerically. We find that the superradiant mode typically experiences three stages: exponential growth, quasiequilibrium, and power-law decay, which can be understood as follows.

Initially, $\epsilon_{211}$ is very small, and its evolution is dominated by the linear term in Eq.~\eqref{nde}, in which case the mode will grow exponentially. The exponential growth continues until the $\gamma_{211}$ becomes almost zero due to black hole spin down in which case the cloud is saturated, or the interaction terms in Eq.~\eqref{nde} become significant. In the former case, the later evolution would be similar to that of free fields. In the later case, the exponential growth could be interrupted by the interaction corrections, and the superradiant cloud reaches a quasiequilibrium stage. By requiring the lhs of Eq.~\eqref{nde} to be zero, one can define the dimensionless occupation number at the quasiequilibrium stage $\epsilon_{eq}$. We expect the cloud reaches the quasi-equilibrium stage before saturation if $\epsilon_{eq} < \epsilon_{max}$.

\begin{figure*}
\includegraphics[height=0.2\linewidth]{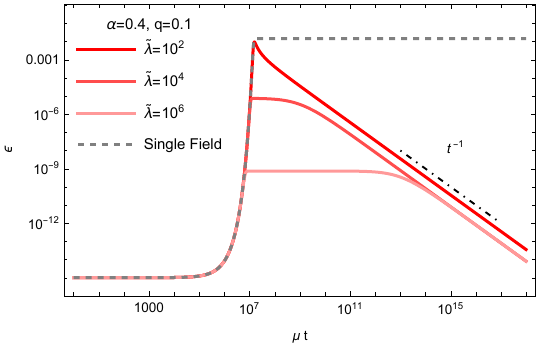}
\includegraphics[height=0.2\linewidth]{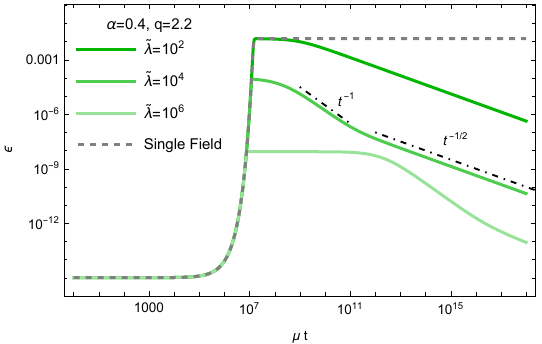}
\includegraphics[height=0.2\linewidth]{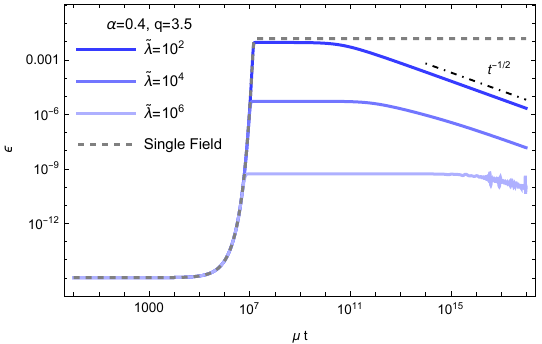} \\
\includegraphics[height=0.2\linewidth]{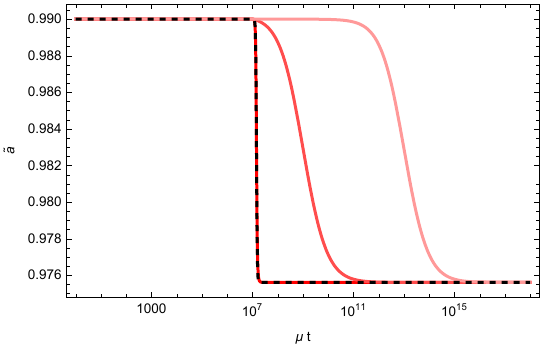}
\includegraphics[height=0.2\linewidth]{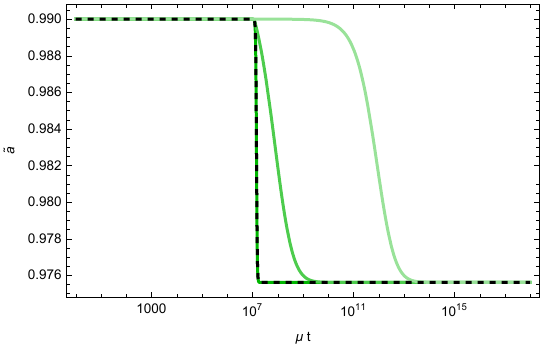}
\includegraphics[height=0.2\linewidth]{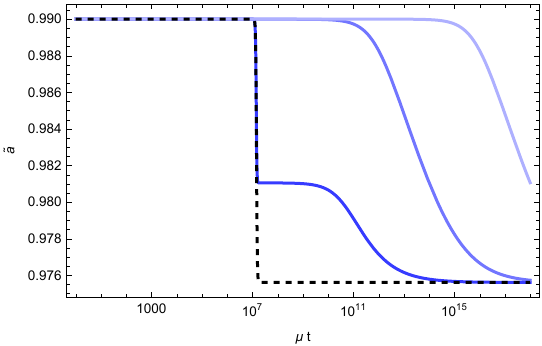}
\caption{Presentative examples of interacting two-field superradiant evolution. The upper panel shows the evolution of occupation numbers in the region of corresponding color in Fig.~\ref{fig:para}, while the lower panel shows the evolution of black hole spin. The black dashed line shows the evolution with single $\bpsi_{211}$ for comparison.}
\label{fig:evo}
\end{figure*}

Although different processes may take place simultaneously as shown in Fig.~\ref{fig:para}, typically only one process dominates the correction, which allows us to simplify the equation by neglecting the subdominating correction. In Fig.~\ref{fig:parad}, we show the parameter space of the dominating process, with the contours denoting the critical $\tilde{\lambda}_c$ defined as $\epsilon_{eq}(\tilde{\lambda}_c) = \epsilon_{max}$. Namely, the cloud would reach the quasiequilibrium stage before it saturates as in the single-field case if $\tilde{\lambda} > \tilde{\lambda}_c$. Here $\tilde{\lambda}$ relates to $\lambda$ via
\be
\lambda = \frac{\tilde{\lambda}}{\sqrt{GM^2}} \mu \approx \left(\frac{ \tilde{\lambda}}{2.7\times 10^{38}}\right) \left( \frac{3 M_\odot}{M} \right) \mu \,.
\ee
Given the value of $\tilde{\lambda}_c$, we find that the cloud reaches the equilibrium stage with very tiny $\epsilon_{eq}$ even if $\lambda / \mu \ll 1$. Depending on the dominating process, the subsequent evolution may be slightly different, shown in Fig.~\ref{fig:evo}, and will be discussed as follows.

\subsection{Case A} \label{sec:caseA}

In region A of Fig.~\ref{fig:parad}, the $\gamma_E$ term in Eq.~\eqref{nde} typically dominates over the other correction terms. In this case, we have $\epsilon_{eq} \simeq \gamma_{211}/\gamma_{E}$, and Eq.~\eqref{nde} and Eq.~\eqref{nda} reduce to
\be
\ba
\epsilon_{211}' &\simeq (\epsilon_{eq} - \epsilon_{211})\epsilon_{211} \\
\ta' &\simeq - \epsilon_{eq} \epsilon_{211} \,,
\ea
\ee
where the prime denotes derivative with respect to the rescaled time $\eta=\gamma_{E}\tilde{t}$. The quasi-equilibrium stage can be reached in a time scale of
\be\label{tausr}
\tau_{sr} = \frac{\ln \epsilon_{eq}- \ln\epsilon_{i}}{\gamma_{211} \mu} \, ,
\ee
where $\epsilon_{i}$ is the initial value of $\epsilon_{211}$ and should be extremely small. Once the quasiequilibrium stage is reached , $\epsilon_{211}$ evolves adiabatically following $\epsilon_{eq}(\ta)$ with the black hole spin satisfying
\be\label{ndaeq}
\ta'=-\epsilon_{eq}^2 \,.
\ee
Without the interruption of other superradiant modes, the quasiequilibrium stage continues until $\epsilon_{eq}$ approaches to $0$ as the black hole spin approaches to the critical spin $\ta_{crit}$. While Eq.~\eqref{ndaeq} can be solved analytically, we shall not write down the lengthy expression of $a$, but are rather interested in the evolution when $\ta$ is close to $\ta_{crit}$, in which case 
\be
\epsilon_{eq} \simeq \frac{\alpha ^{8}}{24 \gamma_E} \left(\frac{1+4\alpha^2}{1-4\alpha^2}\right) (\ta-\ta_{crit}) \,.
\ee
Then Eq.~\eqref{ndaeq} indicates
\be\label{asymp}
\ba
&\ta-\ta_{crit} \simeq \frac{576 \gamma_E}{\alpha ^{16}} \left(\frac{4\alpha^2-1}{4\alpha^2+1}\right)^2 \frac{1}{\tilde{t}}\,, \\
  &\epsilon_{211} \simeq \frac{24}{\alpha^8} \frac{1-4\alpha^2}{1+4\alpha^2} \frac{1}{\tilde{t}} \,,
\ea
\ee
which corresponds the power-law decay as shown in Fig.~\ref{fig:stage}. Note that the other corrections will never dominate the evolution even during the power-law decay. 

\subsection{Case B} \label{sec:caseB}

In region B of Fig.~\ref{fig:parad}, the $\gamma_R$ term typically dominates the interaction correction. In this case, we have
\be
\ba\label{eqcaseb}
\epsilon'_{211} &\simeq \epsilon_{\rm eq}^2 \epsilon_{211} - \epsilon_{211}^3\\
\ta' &\simeq -\epsilon_{\rm eq}^2 \epsilon_{211}.
\ea
\ee
with $\epsilon_{eq} \simeq (\gamma_{211}/\gamma_{R})^{1/2}$. While $\epsilon_{211}$ could also reach $\epsilon_{eq}$ as in case A, the equilibrium may not be always stable. Considering a small deviation $\delta \epsilon$ from $\epsilon_{eq}$ and by perturbing Eq.~\eqref{eqcaseb}, we have
\be
\delta \epsilon' = -2  \epsilon_{\rm eq}^2 \delta \epsilon + \mathcal{O}(\delta \epsilon^2),
\ee
which means a typical time scale for restoring equilibrium is $\tau_{re} \sim 1/{2\epsilon_{\rm eq}^2}$. On the other hand, the $\epsilon_{eq}$ varies on a time scale of
\be
\tau_{var} \sim \left( \frac{\mathrm{d}\epsilon_{\rm eq}}{\mathrm{d}\ta} \epsilon_{\rm eq}^2 \right)^{-1} \,.
\ee
Comparing the two time scales $\tau_{var}$ and $\tau_{re}$ leads to a threshold in terms of black hole spin
\be
\ta_{th} = \frac{\alpha^8}{384\gamma_R} \frac{1+4\alpha^2}{1-4\alpha^2} \,,
\ee
below which the quasiequilibrium stage becomes unstable. If the quasiequilibrium is stable as $a$ approaches $a_{crit}$, the cloud decays with $\epsilon_{211} \propto t^{-1}$. Otherwise, $\epsilon'_{211}$ is dominated by the $\gamma_R \epsilon_{211}^3$ term, and the cloud decays with $\epsilon_{211} \propto t^{-1/2}$.

\subsection{Case C}

In region C of Fig.~\ref{fig:parad}, the $\gamma_B$ term typically first dominates the interaction correction. The evolution is similar to that in case A, except that the black hole spin stops decreasing once reaching the quasiequilibrium stage. The black hole spin and the occupation number remain almost constant until the $\gamma_R$ term becomes significant and takes over the evolution.

\subsection{Cloud collapse}

\begin{figure}
\includegraphics[height=0.6\linewidth]{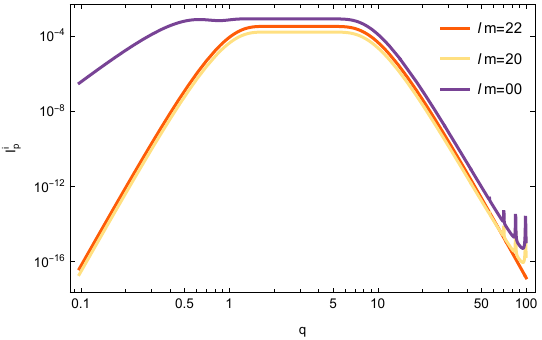}
\caption{Numerical values of each angular momentum component of $I_p(q,\alpha)$. Since the integral is not significantly dependent on $\alpha$, we take $\alpha=0.1$ here as an example.}
\label{fig:IntP}
\end{figure}

The cubic interaction can lead to an effective quartic self-interaction of $\psi$, cf. Fig.~\ref{fig:channels}, under which a cloud may collapse  as the occupation number grows~\cite{Baryakhtar:2020gao,Arvanitaki:2010sy}. The critical occupation number for cloud collapse can be estimated by considering the wave function of $\bar{\psi}_{211}$ with a modified radius $R$
\be
\bar{\psi}_{211} = \frac{\sqrt{N}}{2\sqrt{6}} r R^{-5/2} e^{-r/2R} Y_{11}(\theta,\phi) \, ,
\ee
and investigating how $R$ deviates from the Bohr radius $r_B$ in the presence of the effective self-interaction. In the nonrelativistic limit, the action of $\bpsi$ reduces to 
\be
\ba\label{eq:action}
S_{\bpsi} = \int {\rm d}^3 r {\rm d} t & \frac{i}{2} \left( \bar{\psi}^* \partial_t \bar{\psi} - \bar{\psi} \partial_t \bar{\psi}^* \right) - \frac{1}{2\mu} |\nabla \bar{\psi}|^2 - \mu \Phi_N |\bar{\psi}|^2 \\
& + \bar{\lambda} \bar{\psi}^2 \varphi - \frac{|\nabla \Phi_G|^2}{8\pi G} - \rho_{\rm BH} \Phi_N  \, ,
\ea
\ee
where $\Phi_N$ is the gravitational potential satisfying
\be
\nabla^2 \Phi_N = 4\pi G \left( \rho_{\rm BH}+|\bar{\psi}|^2 \right) \,,
\ee
with $\rho_{\rm BH}=M \delta^3({\bf r})$. $\varphi$ is a mediator. Taking into account the contribution of $s$, $t$, and $u$ channels, we have
\be
\ba \label{eq:propagator}
\varphi = &\left( \sum_{n \ge 3} c_{n} \, v_{n22} + \int {\rm d} k\, c(k) \, v_{k 22} \right) e^{-2i \omega_0 t} \\
& + \left( \sum_{n \ge 3} d_{n} \, v_{n20} + \int {\rm d} k\, d(k) \, v_{k 20} \right) \\
& + \left( \sum_{n \ge 1} e_{n} \, v_{n00} + \int {\rm d} k\, e(k) \, v_{k 00} \right) + c.c. \,,
\ea
\ee
where the coefficients $d_n, d(k),e_n,e(k)$ can be obtained similarly as Eq.~\eqref{eq:coe}. By inserting Eq.~\eqref{eq:propagator} in to action~\eqref{eq:action} and integrating over space, we can obtain an effective potential of $R$, 
\be \label{eq:Veff}
V_{\rm eff} = \frac{\alpha^4 \epsilon}{G \mu} \left( \frac{1}{8\tilde{R}^2} - \frac{1}{4\tilde{R}} - \frac{\tilde{\lambda}^2 \alpha \epsilon I_p}{4\tilde{R}^3} - \frac{711\epsilon \alpha}{1024\tilde{R}} \right) \,,
\ee
where $\tilde{R}=R/r_B$ and $I_p$ is a dimensionless factor containing contribution of $\ell m=22,20,00$, with its numerical values of each component shown in Fig.~\ref{fig:IntP}. The four terms in Eq.~\eqref{eq:Veff} correspond to kinetic energy, black hole gravity, quartic interaction and cloud self-gravity respectively. The extrema of $V_{\rm eff}$ is located at
\be
\tilde{R}_{\rm m}^\pm = \frac{16}{256+711\alpha \epsilon} \left( 8 \pm \sqrt{64-3 I_p \alpha \epsilon \tilde{\lambda}^2 (256+711\alpha \epsilon)} \right).
\ee
However, when the occupation number reaches
\be
\epsilon_{\rm collapse} = \frac{8}{711} \sqrt{\frac{237}{I_p \tilde{\lambda}^2}-256} - \frac{128}{711\alpha}\,,
\ee
one has $\tilde{R}_{\rm m}^+=\tilde{R}_{\rm m}^-$ and the extrema vanishes. In this case, cloud can be unstable, and eventually collapse, leading to a bosenova. When $\tilde{\lambda} \gg 1$, we have $\epsilon_{\rm collapse} \approx (12 I_p \alpha \tilde{\lambda}^2)^{-1}$, which can be very small, indicating the cloud will collapse easily if the coupling is strong.

\section{Implications on Observations}\label{sec:obs}

\begin{figure}
\includegraphics[height=0.6\linewidth]{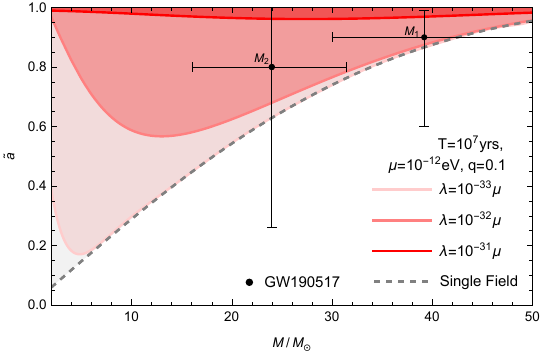}\\
\includegraphics[height=0.6\linewidth]{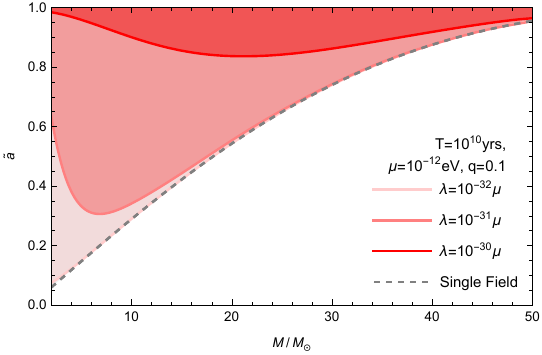}
\caption{Mass-spin distribution of black holes after $10^7$ years (upper panel) and 10 Gyrs (lower panel) superradiant evolution. The red lines show the mass-spin distribution, assuming there are two interacting scalar fields of masses $\mu = 10^{-12}\,\mathrm{eV}$ and $q = 0.1$ with different coupling $\lambda$, while the dashed gray line shows the distribution assuming a single free field. The initial spin of black holes is $a_0 = 0.99$. The crosses in the upper panel show the mass and spin of black holes inferred from GW190517~\cite{LIGOScientific:2021usb}. While the existence of a free scalar field with mass $\sim 10^{-12}~\rm{eV}$ is disfavored by GW190517~\cite{Aswathi:2025nxa}, we find that it could be inconsistent with GW190517 if the field interacts with another field.}
\label{fig:am}
\end{figure}

According to Sec.~\ref{sec:free} and Sec.~\ref{sec:sr}, superradiance evolution may change dramatically in the presence of a second field. In particular, the superradiant efficiency is typically suppressed if the superradiant field couples to another field even with a very weak coupling strength. This effect has important implications on the observational signatures of superradiance, and should be considered in the dark particle searches based on superradiance. For instance, in Fig.~\ref{fig:am}, we show the spin of rapid rotating black holes after $10^7$ yrs and 10 Gyrs, assuming there are two interacting scalar fields with masses $\mu = 10^{-12} \mathrm{eV}$ and $q=0.1$. Depending on the coupling strength, we find the mass-spin distribution can be quite different from the case of a single free field. In particular, the spin measurements from gravitational wave observation GW190517~\cite{LIGOScientific:2021usb} disfavor scalar fields with mass $\sim 10^{-12}~\rm{eV}$~\cite{Aswathi:2025nxa}. Our results indicate that such scalar fields could still be in consistent with the GW190517, if they interact with other fields, cf. Fig.~\ref{fig:am}.

Even with a sufficient growth of the cloud, the existence of the other fields can change the evolution of the cloud, including its maximum mass, decay rate and lifetime, and hence leads to different gravitational wave signals compared to the single-field case. These effects should be taken into account in when searching for dark particles with black hole superradiance.

\section{Conclusion and Discussion}\label{sec:con}

In this work, we investigated black hole superradiance of interacting multiple fields. Taking a two-field model for demonstration, we discuss the possible interactions between the bound and unbound states of the two fields, and their impacts on the superradiant growth via a perturbative approach. We further analyze the superradiant growth of the fastest mode in the presence of the interactions. We find that the superradiant mode typically experiences three stages: the exponential growth stage, the quasiequilibrium stage and the power-law decay stage. We also find that the superradiance is typically suppressed if the superradiant field couples to another field even with a very weak coupling strength. 

While we are taking a two-field model for demonstration, inclusion of more fields should not alter the picture essentially. This is especially the case when there is a large hierarchy in the field mass such that one of the fields grows efficiently, and another field contributes most to the interaction corrections. Nevertheless, one could consider the superradiant growth of the other modes, for example, $\bpsi_{322}$, $\bphi_{211}$ or $\bphi_{322}$ in our setup, which may play a role in the late evolution of $\bpsi_{211}$ in certain parameter regions. Including such modes into the discussion is beyond the scope of this work, and will be studied in future work. It is also interesting to extend the discussion to the spin-1/2 and spin-1 fields, and to consider other possible interactions and their effects on superradiant growth.

It is possible that the dark sector consists of many species of interacting dark particles, rather than just one species of dark particle. As we demonstrated in this work that the existence of the other fields can have important impacts on the observational signatures of superradiance, the superradiance based constraints on dark particles derived from single-field analyses should be revised in the presence of interactions. The observational consequences of multifield black hole superradiance, such as black hole spin distribution, gravitational wave emission from the clouds, and other potential signals of superradiant cloud, are also interesting questions to be studied in future.

\section*{ACKNOWLEDGMENTS}
J. Z. is supported by the National Natural Science Foundation of China (NSFC) under Grants No.~E414660101 and No.~12347103, and the Fundamental Research Funds for the Central Universities under Grants No.~E4EQ6604X2 and No.~E3ER6601A2. Y.-S. P. is supported by National Key Research and Development Program of China under Grant No.2021YFC2203004 and the NSFC under Grant No.12475064.

\appendix

\section{Ultralight scalar fields in the weak gravity limit of the Kerr black hole background}
\label{app:weakgravity}

In this appendix, we shall discuss the solution to the Klein-Gordon equation in the weak gravity limit of the Kerr black hole background. Let us start with the Klein-Gordon equation in the Boyer-Lindquist coordinates
\be\label{eq:KG}
\left(-\Box + \mu^2 \right) \psi = 0 \, .
\ee
We only consider the case that $\alpha \equiv GM \mu \ll 1$, where $M$ is the black hole mass. Far from the black hole horizon, the D'Alembert operator $\Box$ can be expand in powers of $GM/r$. To the first order of $GM/r$, Eq.~\eqref{eq:KG} can be written in the form of
\be
\left[\partial_t^2 - \nabla^2 + \mu^2 - \frac{2GM}{r}\left( \mu^2 + \hat{K}^2 \right) \right]\psi  \simeq 0 \, ,
\ee
where $\nabla^2 = \partial_r^2 + \tfrac2r \partial_r - \tfrac{\hat{L}^2}{r^2}$ is the Laplace operator in Minkowski space and
\be
\ba
\hat{K}^2 &\equiv -\frac{1}{r}\partial_r - \partial_r^2 -\mu^2 - \partial_t^2  \\
&= - \frac{3}{r}\partial_r + \frac{\hat{L}^2}{r^2}  -2  \partial_r^2 
\ea
\ee
with $\hat{L}^2$ being the total angular momentum operator. In the second line of the above equation, we have used the equation of free $\psi$ at zeroth order of $GM/r$ to eliminate the $\partial_t^2$ operator. For nonrelativistic states, the wave number $k$ is much less than $\mu$. Therefore, $\hat{K}^2 \sim k^2$ can be neglected, and we have 
\be\label{eq:nreq}
\left[\partial_t^2 - \nabla^2 + \mu^2 - \frac{2GM\mu^2}{r} \right]\psi  \simeq 0 \, ,
\ee
for nonrelativistic states in the weak gravity limit. 

We shall proceed in the frequency domain by considering
\be
\psi = e^{- i \omega t} u(\mathbf{r}) \,,
\ee
and hence
\be\label{eq:equ}
\left( -\omega^2 + \mu^2 - \nabla^2 - \frac{2 \alpha \mu }{r} \right) u(\mathbf{r}) \simeq 0 \, .
\ee

\be
\left( - \nabla^2 - \frac{2 \alpha \mu }{r} \right) v(\mathbf{r}) \simeq (\sigma^2 - \nu^2) v(r)  \, .
\ee

The general solution to Eq.~\eqref{eq:equ} is well known in the literature. By imposing the regular boundary condition, one can select a set of relevant eigenstates $\{u_{n \ell m}\}$ and $\{u_{k \ell m}\}$, which satisfy
\be
\ba
\left(-\nabla^2 - \frac{2\alpha \mu}{r} \right) u_{n \ell m} &= -k_n^2 u_{n \ell m}, \quad \frac{k_n}{\alpha \mu} = \frac{1}{n}, n \in \mathbb{N} \\
\left(-\nabla^2 - \frac{2\alpha \mu}{r} \right) u_{k \ell m} &= k^2 u_{k \ell m}, \quad \frac{k}{\alpha \mu} \in (0,\infty)
\ea
\ee
respectively. Specifically, the eigenfunctions are given by $u_{n(k) \ell m} = R_{n(k) \ell m} (r) Y_{\ell m} (\theta, \phi)$ with 
\be
\ba
R_{n \ell} =& \sqrt{\left( \frac{2}{n r_B} \right)^3 \frac{(n-\ell-1)!}{2n(n+\ell)!}} e^{-\tfrac{r}{n r_B}} \left( \frac{2r}{n r_B} \right)^\ell L^{2\ell+1}_{n-\ell-1}\left( \frac{2r}{n r_B} \right), \\
R_{k \ell} =& \frac{2k e^{\pi/(2k r_B)} |\Gamma(\ell+1-i/(k r_B))|}{(2\ell+1)!} (2k r)^\ell \\
&\times e^{-i k r} M \left(i/(k r_B)+\ell+1,2\ell+2,2i k r\right) \nonumber
\ea
\ee
where $r_B \equiv 1/\alpha \mu$, $Y_{\ell m}$ is the spherical harmonics, $L^{2\ell+1}_{n-\ell-1}$ is the generalized Laguerre polynomial of degree $n-\ell-1$ and $M(i/(k r_B)+\ell+1,2\ell+2,2i k r)$ is Kummer's function of the first kind. $R_{k \ell}$ are also known as the Coulomb wave function $F_{\ell}(\eta,r)$. Given the asymptotical behavior at spacial infinity, one can find that $\{u_{n \ell m}\}$ correspond to bound states while $\{u_{k \ell m}\}$ correspond to unbound states. The eigenfunctions obey the following orthogonality condition,
\be
\ba
&\int u_{n_1 \ell_1 m_1} u_{n_2 \ell_2 m_2}^* \mathrm{d}^3 \mathbf{r} = \delta_{n_1 n_2} \delta_{\ell_1 \ell_2} \delta_{m_1 m_2} \\
&\int u_{k_1 \ell_1 m_1} u_{k_2 \ell_2 m_2}^* \mathrm{d}^3 \mathbf{r} = 2\pi \delta(k_1-k_2) \delta_{\ell_1 \ell_2} \delta_{m_1 m_2} \\
&\int u_{n \ell_1 m_1} u_{k \ell_2 m_2}^* \mathrm{d}^3 \mathbf{r} = 0 \, .
\ea
\ee

In the literature, the bound states are also often described with $\bpsi$ defined by the nonrelativistic ansatz~\eqref{eq:bpsi}. By substituting ansatz~\eqref{eq:bpsi} into Eq.~\eqref{eq:nreq} and keeping leading order terms in $\alpha$, we can find that $\bpsi$ satisfies
\be
\left(i \partial_t + \frac{\nabla^2}{2\mu} + \frac{\alpha}{r} \right) \bpsi \simeq 0 \, ,
\ee
and the eigenstates are given by
\be
\bpsi_{n\ell m} = e^{- i \xi_{n\ell m} t} u_{n \ell m}(\mathbf{r})
\ee
with $\xi_{n\ell m} \simeq \omega_{n\ell m}-\mu$. For relativistic states, one should take $\hat{K}^2$ into account. Nevertheless, when we calculate the radiation power of the $3\bpsi$ process, we take the flat-space approximation for simplicity, in which case the eigenfunction of the unbound states can be obtained by taking $u_{k\ell m}$ to the limit of $k r_B \rightarrow \infty$, and the radial function $R_{k\ell}$ reduces to the spherical Bessel function.

\section{Green's function method}
\label{app:Green}

In this appendix, we shall show how to solve the inhomogeneous Klein-Gordon equation\footnote{Here we use the $\psi$ field to demonstrate the method, while in the main text, the Green's function method is applied to solve Eq.~\eqref{eq:scphit} and Eq.~\eqref{eq:eomphir} for the $\varphi$ field.}
\be
\left(\Box - \mu^2 \right) \psi(x) = s(x) \,.
\ee
on the Kerr background with the Green's function method. We shall set the Newton's constant $G=1$ in this appendix to avoid confusion with the Green's function. In the Boyer-Lindquist coordinates, the equation is separable with
\be
\psi(x) = \int d\omega e^{-i\omega t} R_{\omega \ell m}(r) S_{\omega \ell m}(\theta) e^{i m \phi} \,
\ee
where $S_{\omega\ell m}(\theta)$ is the spheroidal harmonic function. For simplicity, we shall consider
\be
s(x) = F(r)\, S_{\Omega \ell m}(\theta) \, e^{-i \Omega t} +c.c. \,,
\ee
such that we can focus on the radial function which stratifies
\begin{align}\label{eq:radialTeuEq}
\frac{\mathrm{d}}{\mathrm{d}r}\left(\Delta \frac{\mathrm{d} R}{\mathrm{d}r} \right) +\Bigg[ &\frac{(r^2+a^2)^2\omega^2 -4M a m \omega r + m^2a^2 }{\Delta} \nonumber \\
&- \left(a^2 \omega^2  + \mu^2 r^2  + \Lambda_{\ell m}\right) \left. \Bigg]\right|_{\omega = \Omega}  R=I(r) \,,
\end{align}
where $\Delta=r^2-2Mr+a^2$ and $\Lambda_{\ell m}$ is the eigenvalue of angular Teukolsky equation, and we have suppressed the subscription of the radial function for short.

\subsection{The radial Green's function}

The radial function can be solved with the radial Green's function,
\be\label{eq:RGreen}
R(r_*)=\int_{-\infty}^{+\infty} I(r'_*) \, G_{\ell m \omega=\Omega}(r_*,r'_*) \,\mathrm{d}r'_*
\ee  
where $r_*$ is the tortoise coordinate. The Green's function can be constructed as
\be
G_{\ell m \omega} (r_*,r'_*) = \frac{f_{H}(r_{*<};\omega)f_{+}(r_{*>};\omega)}{\mathcal{W}(f_H,f_+)}
\ee
where $r_{*<} \equiv \min\{r'_*,r_*\}$ ,$r_{*>} \equiv \max\{r'_*,r_*\}$ and $\mathcal{W}(f_H,f_+) \equiv f_H \partial_{r_*} f_+ - f_+ \partial_{r_*} f_H$ is the Wronskian of the two independent homogeneous solutions $f_H$ and $f_+$. The homogeneous solutions are defined by their asymptotic properties 
\be
\ba
f_H &\sim 
    \begin{cases}
        e^{-i k_H r_*} \quad &r_* \rightarrow -\infty \\
        A_\mathrm{in}(\omega) e^{-i k_\infty r_*} + A_\mathrm{out}(\omega) e^{i k_\infty r_*} \quad &r_* \rightarrow +\infty 
    \end{cases} \\
f_+ &\sim 
    \begin{cases}
        B_\mathrm{in}(\omega) e^{-i k_H r_*} + B_\mathrm{out}(\omega) e^{i k_H r_*} \quad &r_* \rightarrow -\infty \\
        e^{i k_\infty r_*} \quad &r_* \rightarrow +\infty 
    \end{cases}
\ea
\ee
where $k_H=\omega-m \Omega_H$, $k_\infty=\sqrt{\omega^2-\mu^2}$. Another useful solution is 
\be
f_- \sim 
    \begin{cases}
        C_\mathrm{in}(\omega) e^{-i k_H r_*} + C_\mathrm{out}(\omega) e^{i k_H r_*} \quad &r_* \rightarrow -\infty \\
        e^{-i k_\infty r_*} \quad &r_* \rightarrow +\infty 
    \end{cases},
\ee
which describes the ingoing wave at infinity. Since there are only two independent solutions, the above solutions are related by
\be
f_H  = A_\mathrm{in}(\omega) f_-  + A_\mathrm{out}(\omega) f_+ \,,
\ee
and the corresponding Wronskians have
\begin{align}
&\mathcal{W}(f_H,f_+) = 2 i k_\infty A_\mathrm{in}(\omega) \,,\\
& \mathcal{W}(f_H,f_-) = -2 i k_\infty A_\mathrm{out}(\omega) \,.
\end{align}

\subsection{Relation to the eigenstates}

\begin{figure*}[htbp]
\includegraphics[height=0.3\linewidth]{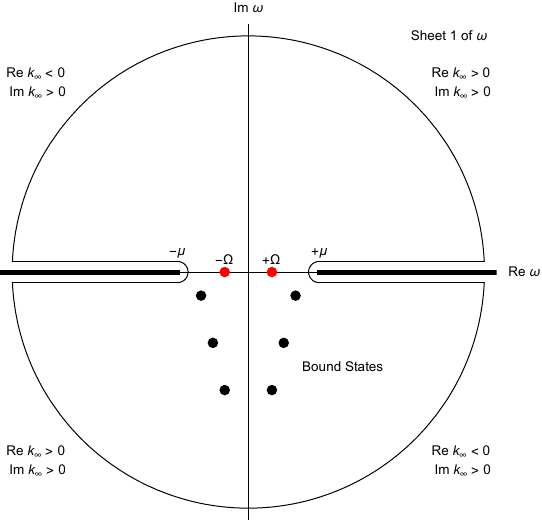}
\quad
\includegraphics[height=0.3\linewidth]{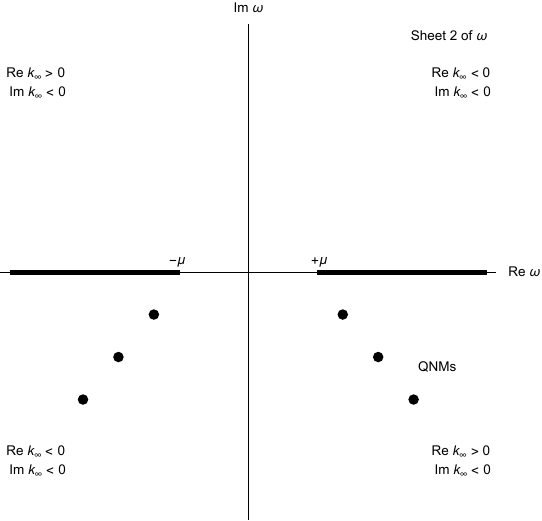}
\caption{Analyticity of $\zeta (z)$ and the integral contour in integral~\eqref{eq:integral}. The branch cut is chosen at $(-\infty,-\mu]$ and $[\mu,\infty)$. The left plot shows the principal Riemann sheet, and the right plot shows the second Riemann sheet. There are poles at $\pm \Omega$ (the red dots), and at the zeros of Wronskian (the black dots). The bound state poles (${\rm Re} \omega < \mu$) are on the principle Riemann sheet, while the qausinormal mode poles are on the second Riemann sheet.}
\label{fig:contour}
\end{figure*}

Now we show that Eq.~\eqref{eq:RGreen} can be expanded with the eigenfunctions of the homogeneous radial equation, i.e., Eq.~\eqref{eq:radialTeuEq} with $I=0$. We shall show this by constructing an integral
\be\label{eq:integral}
\frac{1}{2\pi i} \oint_C \zeta(z)\, \mathrm{d}z \,
\ee
where
\be
\zeta(z) = \frac{2z}{z^2-\Omega^2} \int_{-\infty}^{+\infty} I(r'_*) G_{\ell m z}(r_*,r'_*) \mathrm{d}r'_*\,.
\ee
Recall that $\Omega$ is the forced oscillation frequency of the source term $s(x)$. The integration contour and the analyticity of $\zeta(z)$ are shown in Fig.~\ref{fig:contour}. Because of the multivalue of $k=\pm \sqrt{z^2-\mu^2}$, $\zeta(z)$ has two branch points $z = \pm \mu$ and we define the integral in the Riemann sheet of ${\rm Im}\, k>0$ to fit the outgoing boundary condition, which means the branch cut is $(-\infty,-\mu]$ and $[\mu,\infty)$.

We will see in the following that the integral connects three parts: the residues at $z=\pm\Omega$ which correspond to Eq.~\eqref{eq:RGreen}, the residues at $\omega_{n}$ which correspond to the contribution from bound states, and the branch cut which corresponds to the contribution from scattering states. 

According to the residue theorem, we have
\be
\frac{1}{2\pi i} \oint_C \zeta(z) \mathrm{d}z = \sum_{z_0=\pm \Omega} {\rm Res} \left(\zeta,z_0 \right) + \sum_n {\rm Res} \left(\zeta, \omega_n \right)
\ee
where $\omega_n$ are the relevant poles from $G_{\ell m z}(r_*,r'_*)$. It is straightforward that the residues at $\pm\Omega$ give $R(r_*)$ obtained from Eq.~\eqref{eq:RGreen}. Besides $\pm\Omega$, the zeros of Wronskian also lead to other poles. At these poles, we have $f_H = A_{\rm out} f_+ \propto f_n$, which are the eigenfunctions of bound states and quasi-normal modes by definition. Specifically, the residue can be written
\be
\ba
{\rm Res} \left( \zeta, \omega_n \right) =& \frac{ \int_{-\infty}^{+\infty} I(r'_*) f_{n}(r'_*) \mathrm{d}r'_* }{\omega_n^2-\Omega^2} f_{n}(r_*)\,,
\ea
\ee
where $f_n$ is normalized by
\be
\ba
f_n(r_*) f_n(r_*) \equiv& 2\omega_n {\rm Res} \left( G_z(r_*,r'_*),\omega_n \right) \\
=& -\frac{2\omega_nf_H(r_*;\omega_n) f_H(r_*;\omega_n)}{2i k_\infty A_\mathrm{out} \partial_{\omega_n} A_\mathrm{in}(\omega_n)} \,,
\ea
\ee
such that it approaches the normalized eigenfunctions discussed in Appendix.~\ref{app:weakgravity} in the weak gravity limit.

On the other hand, since $G_{\ell m z}(r_*,r'_*) \sim 1/\sqrt{z^2-\mu^2} $ when $|z| \rightarrow \infty$, the integral going around the circle vanishes, leaving only the contribution from the branch cut. Around the right branch cut, we have
\be
\ba
I_\mathrm{cut} =& \int_\mu^\infty \zeta(z+i \epsilon)-\zeta(z-i \epsilon) \mathrm{d}z \\
=& \int_\mu^\infty \mathrm{d}z \frac{2z}{z^2-\Omega^2} \int_{-\infty}^{+\infty} \mathrm{d}r'_*  I(r'_*) f_H(r_{*<};z) \\
&\times \left( \frac{f_+(r_{*>};z+i\epsilon)}{\mathcal{W}(z+i\epsilon)} - \frac{f_+(r_{*>};z-i\epsilon)}{\mathcal{W}(z-i\epsilon)} \right) \\
=& \int_\mu^\infty \mathrm{d}z \frac{2z}{z^2-\Omega^2} \int_{-\infty}^{+\infty} \mathrm{d}r'_*  I(r'_*) f_H(r_{*<};z) \\
& \times \left( \frac{f_+(r_{*>};z)}{\mathcal{W}[f_H(z),f_+(z)]} - \frac{f_-(r_{*>};z)}{\mathcal{W}[f_H(z),f_-(z)]} \right) \\
=& \int_\mu^\infty \mathrm{d}z \frac{2z}{z^2-\Omega^2} \int_{-\infty}^{+\infty} \mathrm{d}r'_*  I(r'_*) f_H(r_{*<};z) \frac{A_\mathrm{out}f_+ + A_\mathrm{in}f_-}{2i k_\infty A_\mathrm{in}A_\mathrm{out}} \\
=& \int_0^\infty \mathrm{d}k \frac{1}{z^2(k)-\Omega^2} \int_{-\infty}^{+\infty} \mathrm{d}r'_*  I(r'_*) \frac{f_H(r_{*<};z) f_H(r_{*>};z)}{i A_\mathrm{in} A_\mathrm{out}}.
\ea
\ee

Eventually, by equating the above result with the result from the residue theorem, we have
\be
\ba\label{eq:resR}
R (r_*) = & \sum_n \frac{ \int_{-\infty}^{+\infty} I(r'_*) f_{n}(r'_*) \mathrm{d}r'_*}{\omega_n^2-\Omega^2}  f_{n}(r_*) \\
&+ \int_0^\infty \frac{\mathrm{d}k}{2\pi} \frac{\int_{-\infty}^{+\infty} \mathrm{d}r'_*  I(r'_*) f_H(r'_{*};z) }{z^2(k)-\Omega^2} \frac{f_H(r_{*};z)}{A_\mathrm{in} A_\mathrm{out}} \,.
\ea
\ee

For $s(x) = \tilde{s}(\mathbf{r}) e^{- i \Omega t} + c.c.$ with general angular dependence, one can consider the full Green's function~\cite{Yang:2013shb}
\begin{align}
G_{\rm ret}(x,x')=&\frac{2}{ \sqrt{r^2+a^2} \sqrt{r'^2+a^2}} \int d\omega e^{-i\omega(t-t')} \notag \\
&\times \sum_m e^{i m(\phi-\phi')} \sum_l S_{\omega \ell m}(\theta)S_{\omega \ell m}^*(\theta')\notag \\
&\times G_{\ell m \omega}(r,r')
\end{align}
and use Eq.~\eqref{eq:resR} when computing the radial part. Specifically, in the weak gravity limit, we have $A_\mathrm{in} A_\mathrm{out} =1$, $f_n(r) S_{\omega \ell m}(\theta) e^{i m \phi}$ and $f_H(r) S_{\omega \ell m} (\theta) e^{i m \phi}$ reduce to $u_{n \ell m} (\mathbf{r})$ and $u_{k \ell m} (\mathbf{r})$ discussed in Appendix.~\ref{app:weakgravity} respectively. Therefore, the solution to Eq.~\eqref{eq:sourcephi} can be given by Eq.~\eqref{eq:expPhi}.

\section{Stability of quasi-equilibrium}
\label{app:sol}

In Sec.~\ref{sec:caseA} and Sec.~\ref{sec:caseB}, we have discussed the asymptotical behavior near $a_{\rm crit}$ and the condition that the quasiequilibrium becomes unstable. In this appendix, we will show more detailed derivation and generalize our discussion. 

Consider a radiation channel with a $k$-power law, we have
\be
\ba\label{appeq:evo}
\dot{\epsilon} &= \gamma_{sr}(\ta) \epsilon - \gamma_k \epsilon^{k+1} \\
\dot{\ta} &= - \gamma_{sr}(\ta) \epsilon \,,
\ea
\ee
where $k \geq 1, k \in \mathbb{Z}$ and $\gamma_k$ is a constant. Assuming the system is in equilibrium, we can substitute the equilibrium occupation number $\epsilon_{\rm eq}(\ta) = (\gamma_{sr}/\gamma_k)^{1/k}$ into Eq.~\eqref{appeq:evo} and get
\be
\mathrm{d}\eta = - \frac{\mathrm{d}\ta}{\epsilon_{\rm eq}^{k+1}}\,,
\ee
where $\eta=\gamma_{k}\tilde{t}$ is the rescaled time. Integrating both sides of the equation and expand $\ta$ around $\ta_{\rm crit}$, we have
\be
\ba
\eta = k(\ta-\ta_{\rm crit})^{-\tfrac{1}{k}} \left( \frac{\alpha^8}{24\gamma_k} \frac{1+4\alpha^2}{1-4\alpha^2} \right)^{-\tfrac{1+k}{k}} \,,
\ea
\ee
and $\epsilon_{\rm eq}$ can be expanded by
\be
\epsilon_{\rm eq}= \left( \frac{\ta-\ta_{\rm crit}}{\gamma_k} \frac{\alpha^8}{24} \frac{1+4\alpha^2}{1-4\alpha^2} \right)^{\tfrac{1}{k}} \,.
\ee
Hence the occupation number and black hole spin evolve as 
\be
\ba
\ta-\ta_{\rm crit} &= \left( \frac{k}{\eta} \right)^k \left( \frac{24\gamma_k}{\alpha^8} \frac{1-4\alpha^2}{1+4\alpha^2} \right)^{k+1} \\
\epsilon_{\rm eq} &= \frac{24\gamma_k}{\alpha^8} \frac{1-4\alpha^2}{1+4\alpha^2} \frac{k}{\eta} \,.
\ea
\ee
However, the quasiequilibrium stage could be unstable. Considering a small deviation $\delta \epsilon$ from $\epsilon_{\rm eq}$, the evolution of $\delta \epsilon$ is given by
\be
\ba
\delta \epsilon' &= \frac{\gamma_{sr}}{\gamma_k} (\epsilon_{\rm eq}+\delta \epsilon) - (\epsilon_{\rm eq}+\delta \epsilon)^{k+1} \\
&= -k \epsilon_{\rm eq}^k \delta \epsilon \,,
\ea
\ee
which means a typical timescale for restoring equilibrium is $\tau_{re} \sim (k \epsilon_{\rm eq})^{-1}$. On the other hand, $\epsilon_{\rm eq}$ varies on a time scale of 
\be
\tau_{var} \sim \left| \frac{\epsilon_{\rm eq}}{\epsilon'_{\rm eq}} \right| = \left| \frac{\epsilon_{\rm eq}}{\frac{{\rm d}\epsilon_{\rm eq}}{{\rm d}\ta}\ta'} \right| = \left| \frac{\mathrm{d}\epsilon_{\rm eq}}{\mathrm{d}\ta} \epsilon_{\rm eq}^k \right|^{-1} \,.
\ee
Therefore, we expect that equilibrium breaks down when $\tau_{re} \gtrsim \tau_{var}$, as the evolution of real occupation number $\epsilon$ cannot catch up with the change of $\epsilon_{\rm eq}$. In case A, we have $k=1$, and $\tau_{re}<\tau_{var}$ is always satisfied when $a$ is close to $\ta_{\rm crit}$. For $k \geq 1$, there could be threshold in terms of the black hole spin $\ta_{th}$, below which one has $\tau_{re} > \tau_{var}$, indicating the equilibrium is unstable. After the black hole spin drops below $\ta_{th}$, the power-law decay of $\epsilon_{\rm eq}$ transforms from $t^{-1}$ to $t^{-1/k}$.

\bibliography{references}

\end{document}